\documentclass[prl,superscriptaddress,twocolumn,aps,showpacs,amsmath,amssymb,floatfix]{revtex4-1}

\usepackage{color}
\usepackage{bm}
\usepackage{hyperref}
\usepackage{graphicx}
\usepackage{braket}
\usepackage{MnSymbol}
\usepackage{amssymb}
\hypersetup{colorlinks=true}
\renewcommand{\v}[1]{\textbf{\textit{#1}}}
\newcommand{\tr}{\text{tr}}
\newcommand{\Tr}{\text{Tr}}
\newcommand{\KC}{\overset{\circ}{,}}

\newcommand{\diag}[1]{\text{diag}({#1})}

\begin{document}

\title{Crystalline symmetry protected helical Majorana modes in the iron pnictides}

\author{Elio J. K\"onig}
\affiliation{Department of Physics and Astronomy, Center for Materials Theory, Rutgers University, Piscataway, NJ 08854 USA}
\author{Piers Coleman}
\affiliation{Department of Physics and Astronomy, Center for Materials Theory, Rutgers University, Piscataway, NJ 08854 USA}
\affiliation{Department of Physics, Royal Holloway, University of London, Egham, Surrey TW20 0EX, UK}
\date{\today}

\begin{abstract}
We propose that propagating one-dimensional Majorana fermions will 
develop in the vortex cores of certain 
iron-based superconductors, most notably
Li(Fe$_{1-x}$Co$_x$)As. A key ingredient of this proposal
are the 3D
Dirac cones recently observed in ARPES experiments
[P. Zhang et al., Nat. Phys. \textbf{15}, 41 (2019)]. Using an effective Hamiltonian 
around the 
$\Gamma-Z$ line we
demonstrate  the development of gapless one-dimensional helical 
Majorana modes, protected by $C_4$ symmetry. 
A topological index is derived which links the helical Majorana modes 
to the presence of monopoles in the Berry curvature of the 
normal state. We present various 
experimental consequences of this theory and discuss its possible connections
with %quantum information applications and 
cosmic strings. 
\end{abstract}
\date{\today}

\maketitle

Recent
experimental~\cite{ZhangShin2018,ZhangShin2018b,WangGao2018,LiuFeng2018,MachidaTamegai2018,KongDing2019}
and theoretical~\cite{WangFang2015,XuZhang2016,ZhangDasSarma2018}
advances suggest 
that iron-based superconductors (FeSCs) can sustain fractionalized
excitations.  
Building on these ideas, here we propose the 
emergence of dispersive, helical Majorana states in the flux
phase of certain FeSCs.

Twelve years ago, two major discoveries occured in 
condensed matter physics: the observation of high temperature 
superconductivity in the
iron-pnictides~\cite{KamiharaHosono2006,TakahashiHosono2008} and the
discovery of topological insulators
(TIs)~\cite{KoenigMolenkamp2007}. FeSCs have
challenged our understanding of strongly correlated electron
materials, offering the possibility of practical applications. 
%Conventionally, the states near Fermi energy are dominated by the three t$_{2g}$ orbitals of the d-shell, as follows from the crystal field of tetrahedra of ligand
%atoms around the Fe$^{2+}$ ions.
%
%In these layered pnictides or chalcogenides
%Fe$^{2+}$ ions are enclosed in tetrahedral cages of ligand
%atoms. As a consequence of the
%associated crystal field splitting, the electronic states at the
%Fermi energy are dominated by the three t$_{2g}$ orbitals of the d-shell.
%Early magneto-oscillation and
%photoemission experiments corroborated this picture of cylindrical
%Fermi surfaces~\cite{LuShen2008,ColdeaMcDonald2008}.
Topological insulators have transformed our
understanding of band physics 
~\cite{SchnyderLudwig2008, Bernevig2013}
%, leading to the prediction
%of edge, boundary, and surface states in TIs and 
%fully gapped superconductors. 
%Topological field theories and concepts from differential geometry
%suddenly found potential applications to nano-electronics and certain 
%proposals for quantum computation devices. These concepts have 
%been further
and have led to the discovery of 
symmetry protected Weyl and Dirac
semimetals~\cite{ArmitageVishwanath2018}. Remarkably, those materials emulate certain aspects of elementary
particle physics in solid state experiments. 
%In addition, 
%the protected touching points are of potential interest for 
%quantum sensing applications{\color{red}References?}.

%%%%%%%%%%%%%%%%%%%%%%%%%%
\begin{figure}[t]
\includegraphics[width=0.45\textwidth]{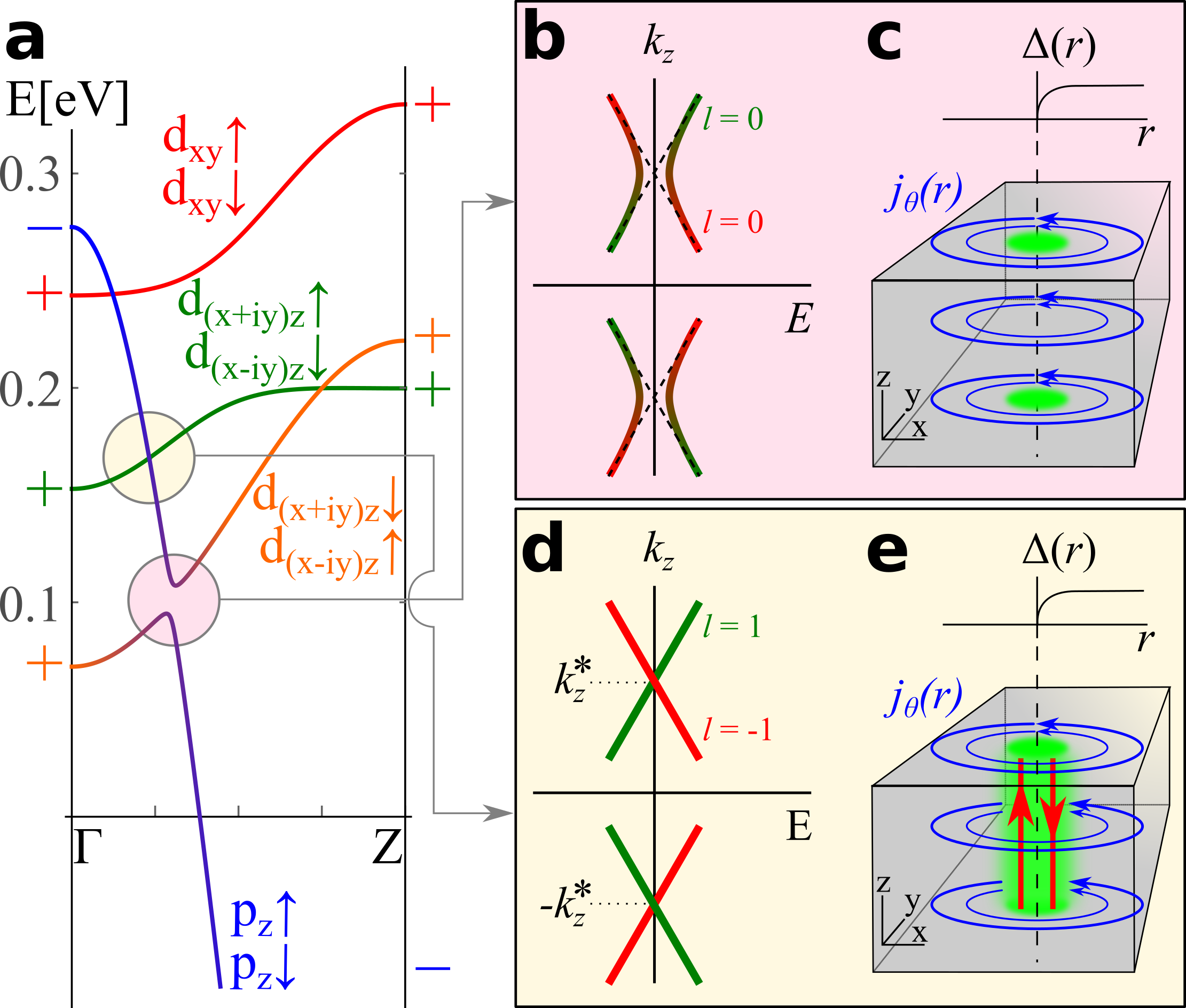} 
\caption{Topology of FeSCs. (a) Band structure in the normal state. For small lattice spacing in \textit{c} direction, $p_z$ orbitals cross the d-states along the $\Gamma$--Z line.
%Spin orbit coupling splits the d-bands. 
When the chemical potential is near the spin-orbit induced gap
(marked by a pink disk around 0.1 eV) the ground state is a
topological superconductor.~(b)~In a vortex core, this implies a
gapped dispersion of bulk Caroli-Matricon-deGennes states with (c)
Majorana zero modes (green pancakes) at the surface termination.
At higher doping, when the Fermi energy 
lies in the vicinity of the 
Dirac node (marked by a yellow disk around 0.17
eV), (d-e) $C_4$ symmetry protects helical Majorana states dispersing along the vortex cores. }
\label{fig:Dispersions}
\end{figure}
%%%%%%%%%%%%%%%%%%%%%%%%%%

Yet despite the excitement in these two new fields, 
until recently, there has been little overlap between them. 
Iron based superconductors are layered structures, 
in which d-orbitals of the iron atoms
form quasi-two dimensional bands. 
%atoms around the Fe$^{2+}$ ions.
The spin-orbit coupling (SOC) in the d-bands 
was long thought to be too small for 
topological behavior.  However, the recent
discovery of marked spin-orbit splitting 
in photoemission spectra~\cite{JohnsonFederov2015,BorisenkoZhigadlo2016} has overturned
this assumption, with an observation 
~\cite{WangFang2015,BorisenkoZhigadlo2016,XuZhang2016,ZhangShin2018}
that at small interlayer separations, 
an enhanced c-axis dispersion drives 
a topological band inversion between the iron d-bands and ligand
$p_{z}$ orbitals. 
%leads to a band-crossing, driving the corresponding FeSC 
%into a topological state.
When the chemical potential lies in the hybridization gap between the
d and p-bands, 
%the band crossing 
%this implies that 
the corresponding topological FeSCs sustain Majorana zero modes
where-ever magnetic flux lines intersect with the surface,
Fig.~\ref{fig:Dispersions} (c).
These excitations have been observed~\cite{WangGao2018,LiuFeng2018,MachidaTamegai2018,KongDing2019}.
Here we 
demonstrate that on additional doping, topological behavior
is expected to give rise to dispersive, helical Majorana
fermions, Fig.~\ref{fig:Dispersions} (e) along the cores of
superconducting vortices. Observation of these excitations
%of the latter 
would provide an important confirmation of the topological character of
iron based superconductors, yielding a new setting for the
realization of Majorana fermions. 

Helical Majorana fermions in one dimension correspond to a pair of gapless
counter-propagating fermionic excitations, first proposed 
as excitations within the ``$o-$'' vortices of 
superfluid $^3$He-B ~\cite{MisirpashaevVolovik1995}. 
%To date however, 
Whilst these excitations have not been observed, 
possibly because the energetics of $^3$He-B favors less
symmetric $v-$vortices \cite{VollhardtWoelfleBook,VolovikBook}, which 
do not support helical Majorana modes, 
we here propose an alternative realization in FeSCs.
Recent experimental advances (Table ~\ref{tab:1DMajo}) provide evidence
for chiral (i.e. unidirectional) Majorana modes at the boundaries of
various two dimensional systems, including 
superconducting-quantum anomalous Hall
heterostructures \cite{HeWang2017}, 5/2 fractional quantum Hall
states\cite{BanerjeeHeiblum2018} and the layered
Kitaev material $\alpha$-RuCl$_3$~\cite{KasaharaMatsuda2018}.

\begin{table*}
\begin{tabular}{|r||r l||r l|}
\hline
 & &boundary of 2D systems & & vortex in 3D system \\ 
\hline \hline
chiral & \begin{tabular}{r|}
Exp. \\
\hline
Th. 
\end{tabular} &\begin{tabular}{l}
QAH-SC \cite{HeWang2017}, $\alpha$-RuCl$_3$ \cite{KasaharaMatsuda2018}, $\nu = 5/2$ QH \cite{BanerjeeHeiblum2018} \\
\hline
$p+ip$ SC~\cite{Volovik1988} (Sr$_2$RuO$_4$ \cite{RiceSigrist1995}?) [AZ cl. D \cite{SchnyderLudwig2008}]
\end{tabular} &  \begin{tabular}{r|}
Exp. \\
\hline
Th. 
\end{tabular} &\begin{tabular}{l}
N/A\\
\hline
TI-SC heterostructure~\cite{MengBalents2012}[Weyl SSM]
\end{tabular}  \\ 
\hline \hline
helical & \begin{tabular}{r|}
Exp. \\
\hline
Th. 
\end{tabular} & \begin{tabular}{l}
 N/A\\
\hline
NCS \cite{TanakaNagaosa2009, SatoFujimoto2009,Roy2008,QiZhang2009}, $s_\pm$ SC+SOC~\cite{ZhangMele2013} [AZ cl. DIII \cite{SchnyderLudwig2008}]
\end{tabular} & \begin{tabular}{r|}
Exp. \\
\hline
Th. 
\end{tabular} & \begin{tabular}{l}
 N/A \\
\hline
 $^3$He-B \cite{MisirpashaevVolovik1995}, LiFe$_{1-x}$Co$_{x}$As (this work) [Dirac SSM]
\end{tabular} \\
 \hline
\end{tabular} 
\caption{Phases of matter which sustain 1+1D helical or chiral Majorana fermions. We present all experimental evidence, the first material specific theoretical proposal and generic classes of systems (in square brackets).  We omitted Majorana modes which occur at fine-tuned critical points, e.g. at topological phase transitions~\cite{HosurVishwanath2011, XuZhang2016} or at S-TI-S junctions with flux $\pi$~ \cite{FuKane2008}. Abbreviations: ``AZ cl.'' = ``Altland-Zirnbauer class'', ``Exp.'' = ``Experiment'', ``NCS'' = ``non-centrosymmetric superconductor'', ``QAH'' = ``Quantum anomalous Hall'', ``QH'' = ``Quantum Hall'', ``SC''=``superconductor'', ``SSM'' = ``superconducting semimetal'', ``Th.'' = ``Theory''.
}
\label{tab:1DMajo}
\end{table*}

\textit{Majorana modes in FeSC.}  Here we summarize the main physics
leading to the appearance of helical Majorana subgap states in the
flux phase of FeSC, when the magnetic field is aligned in \textit{c}
direction. 
%Mathematical details are contained in \cite{SuppMat}, while
%the experimentally realistic case of misaligned magnetic field and
%crystal axis is discussed below. 
We shall concentrate on a case where
the vortex core size, (determined by the coherence length), is much
larger than the lattice spacing, so that vortex-induced interpocket
scattering can be neglected. This permits us to concentrate on the region
of the Brillouin zone (BZ) which harbors the topological physics, in this
case the $\Gamma-Z$ line.

Along this line, the relevant electronic states are classified by
the $z-$component of their total angular momentum $J_{z} = L_{z} +
S_{z}$. We may exploit the fact that the low energy 
%$\v k \cdot \v p$ 
Hamiltonian
close to the $\Gamma-Z$ line~\cite{XuZhang2016,ZhangShin2018,SuppMat}
features an emergent continuous rotation symmetry. 
Three pairs of states are important, $\ket{d_{(x+iy)z}
\downarrow}, \ket{d_{(x-iy)z} \uparrow}$ (with $j_z = \pm 1/2$),
$\ket{p_z,\uparrow}, \ket{p_z,\downarrow}$ (also $j_z = \pm 1/2$) and
$\ket{d_{(x+iy)z} \uparrow}, \ket{d_{(x-iy)z} \downarrow}$ ($j_z = \pm
3/2$). Their dispersion is shown in Fig.~\ref{fig:Dispersions} along
with the $d_{xy}$ bands, we used the low energy model of
Ref.~\cite{XuZhang2016}.

We briefly recapitulate the appearance of localized Majorana zero
modes. The $j_{z}= \pm 1/2$ $p_z$ states can hybridize
with the corresponding $\ket{d_{(x+iy)z} \downarrow},
\ket{d_{(x-iy)z} \uparrow}$ states at intermediate $k_z$, leading to an
avoided crossing of the bands 
[pink circle at 0.1 eV in
Fig.~\ref{fig:Dispersions} a)]. Since the $p$ and $d$ orbitals  carry
opposite parity, the band-crossing 
leads to a parity inversion at the 
the Z-point. 
The system is therefore topological~\cite{FuKane2007}.
%,  in the
%sense that, if the Fermi energy, shifted into the gap at 0.1
%eV, did not
%cross any other bands, the system would be a 3D TI. 
In the superconducting state, this system 
is then expected~\cite{FuKane2008} to host topological surface
superconductivity, developing  localized Majorana zero modes at the
surface termination of a vortex,
Fig.~\ref{fig:Dispersions} c).
These Majorana zero modes can be alternatively
interpreted as the topological end states of a fully gapped, 1D
superconductor inside the vortex core~\cite{XuZhang2016}. In
the bulk,  where $k_z$ is a good quantum number the vortex
hosts fermionic subgap states for each $k_z$ near the normal state
Fermi surface, Fig.~\ref{fig:Dispersions} b). In particular, the lowest lying states carry angular momentum $l =
0$ and develop a topological hybridization gap upon inclusion
of SOC.
%in the
%absence of SOC, two gapless helical bulk Majorana
%branches develop; since they carry angular momentum $l =
%0$, a topological hybridization gap opens upon inclusion
%of SOC.

However, bulk FeSCs can also support dispersive 
helical Majorana modes in their vortex cores. To see this, 
we now turn to the situation where the chemical potential
lies near the Dirac cone, highlighted by a yellow circle at about 0.17 eV in
Fig.~\ref{fig:Dispersions} a). At this energy, semimetallic Dirac
states are observed in ARPES~\cite{ZhangShin2018}: these 
occur because the different $j_z$ quantum numbers of
$\ket{p_z,\uparrow}, \ket{p_z,\downarrow}$ and $\ket{d_{(x+iy)z}
\uparrow}, \ket{d_{(x-iy)z} \downarrow}$ prevent a hybridization on
the high symmetry line leading to a Hamiltonian of the (tilted) Dirac
form $H(\v k) = H_+(\v k) \oplus H_-(\v k)$
\cite{ZhangShin2018,SuppMat},
%%%%%%%%%%%%%%%%%%%
\begin{equation}\label{eq:DiracHam}
H_\pm(\v k) = \left (\begin{array}{cc}
M_{p}(k_z) & \pm v k_x + i v k_y \\ 
\pm v k_x - i v k_y & M_{d}(k_z)
\end{array} \right ) ,
\end{equation}
%%%%%%%%%%%%%%%%%%%
where $H_+(\v k)$ ($H_-(\v k)$) acts in the subspace of positive (negative) helicity spanned by $\ket{p_z,\uparrow}, \ket{d_{(x+iy)z} \uparrow}$ $(\ket{p_z,\downarrow},\ket{d_{(x-iy)z} \downarrow})$. The dispersion $M_{p}(k_z),M_{d}(k_z)$ of the relevant $p$ and $d$ orbitals is plotted in Fig.~\ref{fig:Dispersions} a) and $v$ is the transverse velocity.

We now assume that below $T_{c}$, 
a spin-singlet, 
s-wave
superconducting phase develops.
%with significant intraorbital pairing, develops. 
In an
Abrikosov lattice of vortex lines, translational symmetry allows to
solve the problem at each $k_z$ separately. At the
particular values of $k_{z}= \pm k_z^*$, where
$M_p(\pm k_z^*) = M_d(\pm k_z^*)$,  $H_+$ and $H_-$ 
separately take the form of a TI surface state. 
Consequently~\cite{FuKane2008}, for each helicity a non-degenerate
Majorana zero mode appears in each vortex. 
Now in contrast to the case of Fig.~\ref{fig:Dispersions} b), these two modes carry different angular momenta $l = \pm 1$ so that
they can not be mixed by any perturbation which respects the $C_4$
symmetry. This leads to the gapless linear helical dispersion near $\pm k_z^*$.

\begin{figure}[t]
\includegraphics[width=0.47\textwidth]{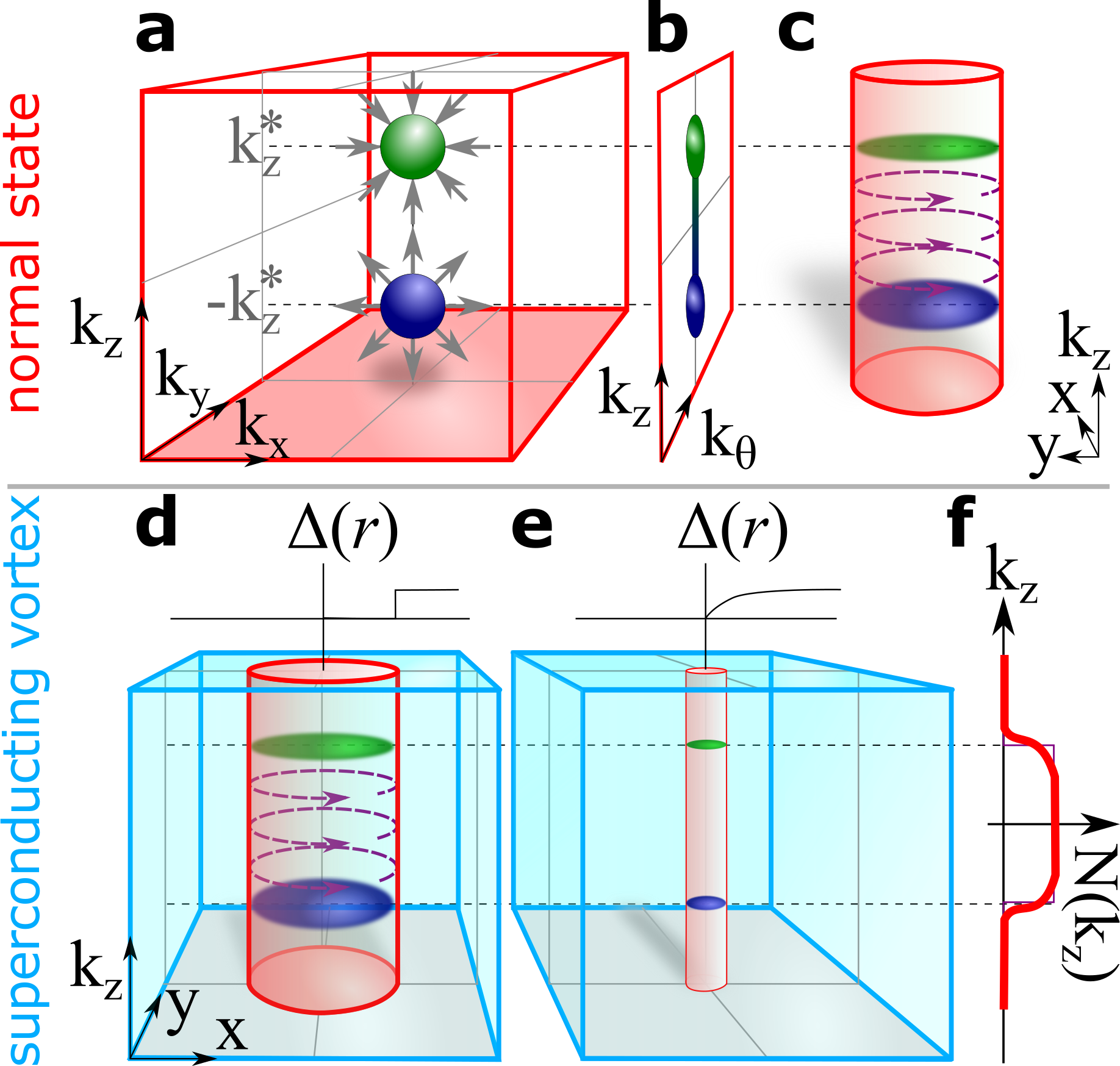} 
\caption{Topological origin of helical Majorana modes. (a) Normal state BZ with helicity resolved Weyl nodes. (b) Surface BZ and Fermi arc. (c) Partial real space representation of a cylindrical Weyl semimetal. Fermi arc states (purple circles) terminate at topological transitions at $\pm k_z^*$ with delocalized, critical bulk states (green and blue pancakes). (d) A fat vortex: a normal state cylindrical core (red) embedded in a fully gapped superconductor (light blue). (e) A realistic thin vortex: Fermi arc states are finite size gapped, but the $\pi$ Berry phase at $\pm k_z^*$ protects the critical states in the core~\cite{FuKane2008,SuppMat}. (f) The index $N(k_z)$, Eq.~\eqref{eq:IndexMainText} (purple, thin), which we semiclassically relate to $\sigma_{xy}(k_z)$, Eq.~\eqref{eq:IndexSemiclassical} (red, thick).}
%
%\caption{(a) As the crystal momentum $k_z$ is varied, the Dirac nodes in the normal state bulk Brillouin-zone imply a family of quantum spin Hall states for each momentum $k_z \in (-k_z^*,k_z^*)$ (here Weyl nodes in the sector of positive helicity are shown). The topological phase transition at $k_z = \pm k_z^*$ imposes the existence of critical bulk states, which coincide with the termination points of the Fermi arc in the surface BZ (b). (c) A superconducting vortex consists of a metallic core enclosed by a fully gapped superconductor which formally leads to Majorana Fermi arc states (purple, dashed) encircling the core. (d) While such states are typically gapped out for realistically small core diameter $\xi$, the change in topology at $\pm k_z^*$ enforces the spectral flow of states presented in Fig.~\ref{fig:Dispersions} (d). (d, inset) This is encoded in the index $N_+(k_z)$, Eq.~\eqref{eq:IndexMainText} (purple solid), which we semiclassically relate to the Hall response, Eq.~\eqref{eq:IndexSemiclassical} (red, dashed).
%When $\mu \neq 0$, $\sigma_{xy,+}(k_z)$ is smeared near $\pm k_z^*$.}
\label{fig:Origin}
\end{figure}

\textit{Topological origin of helical Majorana modes.}  The
crystalline topological protection of the helical Majorana modes in
the flux phase of FeSC can be understood as follows. First, we
note that in the normal state, crystalline symmetries, in
particular $C_4$, impose the decoupling of Hamiltonian
\eqref{eq:DiracHam} into the direct sum of two decoupled helical
sectors. Within $H_+$ ($H_-$), two Weyl points of opposite topological
charge $\pm 1$ ($\mp 1$) appear at $(0,0,\pm k_z^*)$,
Fig.~\ref{fig:Origin} a). Since crystalline symmetry ensures perfect
decoupling, it is favorable to concentrate on a given sector in these
explanations and superimpose both sectors in the end. The Berry flux
connecting the two Weyl points implies a quantum anomalous Hall state
for $k_z \in (-k_z^*, k_z^*)$~\cite{FootnoteFS}. The resulting family
of chiral edge states forms a Fermi arc in the surface BZ, Fig.~\ref{fig:Origin} b,c). In view of their chiral
nature, Fermi arc states can only terminate at a $k_z$ which sustains critical
bulk states - i.e.~at the projection of the Weyl points. From the boundary perspective, their presence is ensured by the topological phase transition at $\pm k_z^*$.

%The presence
%of these critical bulk states is ensured since the %system undergoes a
%topological phase transition as a function of $k_z$.

We now turn to the superconducting case in the flux phase, for which a
vortex core represents a normal state cylinder inside of a fully
gapped superconducting background. At each $k_z \in (-k_z^*, k_z^*)$
the boundary of the vortex core resembles an interface between
quantum anomalous Hall state and topological superconductor.  This 
leads to a chiral Majorana encircling the cylinder - %this is 
i.e.~the
Majorana analog~\cite{ZhangShin2018,YangZhang2014} of Fermi arc states (purple
circles, Fig.~\ref{fig:Origin}, d). 
%At the same time, for $k_z \notin
%[-k_z^*, k_z^*]$ Fermi arc states are absent. 
As explained above, edge
states may only disappear as a function of $k_z$ when the bulk is
critical, therefore it follows that topologically protected vortex
core subgap states must cross the Fermi energy at $\pm k_z^*$.

We conclude this discussion with three remarks. (1) For typical vortex
core diameters $\xi$ the chiral Majorana edge states are gapped by finite size
effects, yet the above topological argument is still valid, Fig.~\ref{fig:Origin} e). 
In particular, as
in the case of a 3D TI surface, the magnetic flux prevents 
the critical bulk (= vortex core) states at $\pm k_z^*$ from gapping. A different situation occurs in $^3$He-A, where the conservation of the spin projection protects the non-dispersive Fermi arc states for all $k_z$ between the projection of Weyl points~\cite{VolovikBook, Volovik2011, SuppMat}. (2)
Taking into account that $H_+$ and $H_-$ sectors have opposite
helicity, the actual state for $k_z \in (-k_z^*, k_z^*)$ is a quantum
spin Hall insulator, and Fermi arc states are helical rather than
chiral. (3) For weak misalignment of the flux line and the \textit{c}-axis, mixing between decoupled helical sectors $H_\pm$ is negligible. Under this assumption, the topological protection of helical modes persists.

\textit{Bogoliubov-deGennes (BdG) Hamiltonian.} To confirm these heuristic arguments, we have perturbatively diagonalized~\cite{SuppMat} the BdG Hamiltonian of a topological FeSC with a single vortex. Here, we concentrate on states near $k_z^*$ and employ a simplified Hamiltonian 
$\mathcal H = \mathcal H_{+}\oplus \mathcal{H}_{-}$,
where  
\begin{eqnarray}\label{l}
\mathcal H_\pm = (H_\pm - \mu) \tau_z + \Delta (\v r) \tau_{+}
+\Delta^* (\v r) \tau_{-}.
\end{eqnarray}
The Fermi
energy $\mu$ is measured from the Dirac point, $\Delta (\v r) = \vert
\Delta(r) \vert e^{i \theta}$ is the superconducting gap ($\vert
\Delta(\infty) \vert \equiv \Delta$) and $\tau_{x,y,z}$ are Pauli
matrices in Nambu space and $\tau_{\pm}= (\tau_{x}\pm i \tau_{y})/2$.
Assuming circularly symmetric vortices, we expand the
wave function in angular momenta, seeking solutions of the form
$ \Psi_\pm =
\sum_{l} e^{i k_z z + i l \theta} U_\pm(\theta) \Psi_\pm^{(l)}(r,k_z) $,
where the precise form of the diagonal matrices $U_\pm(\theta)$ is given in the supplement. 
%$U_+ = \diag{u,u_{(l-1)}}$, $U_-^{(l)} =
%\diag{u^*_{(-l-1)},u^*_{(-l)}}$ and $u_{(l)} = (e^{i l \theta - i
%\pi/4}, e^{i (l-1) \theta + i \pi/4})$.  
At $l = \pm 1$ a chiral
symmetry in the $l$-th sector $\mathcal H_\pm^{(l)}$ allows us to
explicitly construct an unpaired zero energy solution $\Psi_\pm^{(\pm
1)} (r,k_z^*)$ in each helical sector.
%= (v_\pm(r), \pm v_{\mp}(r))$, where $v_\pm(r) =
%\mathcal N e^{- \int_0^r \Delta(r') dr'/v} (\mp J_{1/2\pm1/2}(\mu
%r/v),J_{1/2\mp1/2}(\mu r/v))$, $\mathcal N$ is a normalization
%constant and $J_\nu(x)$ are Bessel functions.  
We use these solutions
to perturbatively include momenta $k_z - k_z^*$, a Zeeman field
$g \mu_B B/2$ and orbital dependent gaps $\Delta_p - \Delta_d = \delta \Delta
\neq 0$. By projecting onto the low-energy space we obtain the effective dispersions 
\begin{equation}
E_\pm(k_z) = \pm [v_M (k_z - k_z^*)  -w_M \delta \Delta + g \mu_B B/2].
\end{equation}
%\begin{equation}
%\mathcal H_{\rm eff} = \bigoplus_\pm \mp[v_M p_z + w_M\delta \Delta - E_Z]	,
%\end{equation}
%{\color{red}  What does this mean?  Where are the operators? 
%}
This confirms the heuristic argument for the appearance of helical Majorana modes and demonstrates that perturbations merely shift $k_z^*$.  A similar result holds near $-k_z^*$, so that in total two pairs of helical Majorana modes occur, Fig.~\ref{fig:Dispersions} d). In the limit $\mu \gg \Delta$ we obtain $v_M
\sim \Delta^2 \partial_{k_z^*}[M_d(k_z^*) -M_p(k_z^*)]/\mu^2$ and $w_M \sim \Delta/\mu$. The velocity
of helical Majorana modes in vortices of $^3$He - B has an analogous
parametrical dependence~\cite{MisirpashaevVolovik1995}. 
%
%
%inclusion~\cite{SuppMat} of momenta near $k_z^*$ reveals a linear
%helical dispersion, Fig.~\ref{fig:Dispersions} d), with velocity $v_M
%\sim \Delta^2 [M_p'(k_z^*) -M_d'(k_z^*)]/\mu^2$, where $\Delta$ is the
%superconducting gap, $\mu \gg \Delta$ the Fermi energy measured from
%the Dirac node and the prime denotes a $k_z$ derivative. The velocity
%of helical Majorana modes in vortices of $^3$He - B has an analogous
%parametrical dependence~\cite{MisirpashaevVolovik1995}.

\textit{Index theorem.} We now demonstrate the link between the
helical Majorana modes and the Berry flux between the
two pairs of Weyl points, Fig.~\ref{fig:Origin}. While several topological invariants were proposed~\cite{Weinberg1981,
TeoKane2010,QiZhang2013,RoyGoswami2014} to describe dispersive Majorana modes, we here
employ a generalization of an index introduced by
Volovik~\cite{Volovik1989} for vortices in $^3$He. 
Our index measures the imbalance between the number of states
of opposite helicity at a given momentum $k_{z}$, 
$N(k_z) = [N_-(k_z)-
N_+(k_z)]/2$, 
where 
%\begin{equation}
%N_{\pm} (k_z) = \text{Im} \int_{-\infty}^0 \frac{d\omega}{\pi}\Tr[\mathcal G_\pm(\omega-i0)]e^{\omega 0^+}, \label{eq:IndexMainText}
%\end{equation}
\begin{equation}
N_{\pm} (k_z) = \sum_{n} \theta[-E_n^\pm(k_z)] = \text{Im} \int_{-\infty}^0 \frac{d\omega}{\pi}\Tr[\mathcal G_\pm(\omega-i0)]e^{\omega 0^+} \label{eq:IndexMainText}
\end{equation}
counts the number of states with given helicity in the Fermi sea ($\mathcal G_\pm(z) = [z - \mathcal H_\pm (k_z)]^{-1}$ and $n$ labels quantum numbers).
% with quantum number $n$. 
In a fully gapped system $N(k_z)$ is constant as a function of $k_z$. In contrast, the presence of helical Majorana modes, Fig.~\ref{fig:Dispersions} d), implies a jump $N(k_z^*+0^+) - N(k_z^*-0^-) = 1$. 

We now relate $N(k_z)$ to the quantized spin Hall conductivity at a
given $k_z$ using a semiclassical expansion, which is valid for
smoothly varying $\Delta(\v r)$. In the eigenbasis of the normal state
Hamiltonian $H_\pm({\v p}) \ket{u_{\v p, \xi,\pm}} = \epsilon_{\xi,
\pm}(\v p) \ket{u_{\v p, \xi, \pm}}$ the BdG Hamiltonian in each band
takes the form $\mathcal H_{\xi,\pm} = \v d_{\xi,\pm} \cdot
\boldsymbol \tau$ (where the transverse components of $\v d$ describe
intraorbital pairing). In the following argument, we 
% concentrate on $N_+(k_z)$ and
drop the band and helicity indices $\xi$ and $\pm$ and employ a Wigner transform~\cite{SuppMat,KoenigLevchenkoUP} so that $\v d(\v
R, \v P) = (\text{Re} \Delta(\v R), - \text{Im} \Delta(\v R),
\epsilon(\v P) - \mu)$.
%Semiclassically $\v d(\v R, \v P) = (\text{Re} \Delta(\v R), - \text{Im} \Delta(\v R), \epsilon(\v P) - \mu)$, which may be obtained by means of a Wigner transform which explicitly respects gauge invariance $\ket{u_{\v p}} \sim e^{i \phi(\v p)} \ket{u_{\v p}}$~\cite{SuppMat,KoenigLevchenkoUP}. 
Due to the algebra of Pauli matrices, the Green's function $\mathcal
G(\omega; \v R, \v P) = [\omega - \v d \cdot \boldsymbol \tau]^{-1}$
contains a commutator of operator convolutions (denoted by $\circ$)
\begin{equation}
\mathcal G(\omega; \v R, \v P)= (\omega + \v d \cdot \tau) \circ [\omega^2 - \v d^2 - \frac{i}{2} \epsilon_{abc} [d_a \KC d_b] \tau_c]^{-1}. \label{eq:GreensMainText}
\end{equation}
The gradient expansion of the convolution is
\begin{align}
[d_a \KC d_b](\v R, \v P) & \simeq i\left (\vec \nabla_X d_a \cdot \vec \nabla_P d_b - \vec \nabla_P d_a \cdot \vec \nabla_X d_b\right ) \notag \\
& + i \Omega_z \hat e_z\cdot \left (\vec \nabla_X d_a \times \vec \nabla_X d_b\right ). \label{eq:MoyalMainText}
\end{align}
Here, $\Omega_z = i  \braket{\partial_{p_x} u_{\v p} \vert \partial_{p_y} u_{\v p}} - i \braket{\partial_{p_y} u_{\v p} \vert \partial_{p_x} u_{\v p}}$ is the Berry curvature. 
Note that within our gauge invariant formalism, the semiclassical coordinates $\v R, \v P$ are \textit{kinematic} -- this leads to the appearance of $\Omega_z$ in addition to the Poisson bracket~\cite{Altshuler1978}.
%In complete analogy, a magnetic field
%appears in the Moyal expansion 
%when the Wigner transform is defined in a manner to preserve gauge invariance in real space $\psi(\v x) \sim e^{i \phi(\v x)} \psi(\v x)$

We evaluate $N(k_z)$ for an isotropic vortex of winding $\nu_v$ to leading order in gradients. The vortex enters Eq.~\eqref{eq:MoyalMainText} as $\vec \nabla_X d_y \times \vec \nabla_X d_x = \nu_v \hat e_z [\partial_R \vert \Delta(R) \vert^2]/2R$. Performing the radial 
%We perform the $R$ 
integration 
%for a finite system 
and restoring the band and helicity indices, leads to the 
result $N_\pm(k_z) = - \nu_v \sigma_{xy,\pm}(k_z)$, where
%%%%%%%%%%%%%
\begin{equation}
 \sigma_{xy}^{\pm}(k_z) = \sum_\xi \int \frac{dk_{x}dk_{y}}{2\pi}
 \Omega^{\xi,\pm}_z (\v k) \theta(-\epsilon_{\xi,\pm}(\v k )). \label{eq:IndexSemiclassical}
\end{equation}
%%%%%%%%%%%%%
In this expression, $\Omega^{\xi,\pm}_{z} (\v k)$ and 
$\epsilon_{\xi,\pm}$ are evaluated in the plane at constant $k_z$. It  follows that $N(k_z) = \nu_v [\sigma^{+}_{xy}(k_z)-\sigma^{-}_{xy}(k_z)]/2$ is given by the normal state spin Hall conductivity 
%at momentum $k_z$ 
which establishes
%semiclassically proves 
the topological origin of the jump in the
Fermi surface volume, Fig.~\ref{fig:Origin} f). 
%%%%%%%%%%%%%%%%%%%%%%%%%%

\textit{Experimental realization.}  
%It is important to stress that the
%non-interacting topological picture which we employed relies on
%ab-initio density functional theory (DFT) with limited reliability in
%the regime of realistic substantial interactions.  In order to provide
%support for the topological paradigm associated with
%Fig.~\ref{fig:Dispersions}, w
We now summarize the topological
features of iron-based superconductors observed to
date. 
%Firstly, t
Topological Dirac surface states have been detected in
Fe(Te$_{x}$Se$_{1-x}$) and Li(Fe$_{1-x}$Co$_x$)As using (S)ARPES, both
in the normal and superconducting
states~\cite{ZhangShin2018b}, while photoemission evidence for 3D
Dirac semimetallic bulk states in the normal state was also reported
in Ref.~\cite{ZhangShin2018}. Moreover, zero bias peaks in vortices of
the flux phase of Fe(Te$_{x}$Se$_{1-x}$)~\cite{WangGao2018,MachidaTamegai2018,KongDing2019} and
(Li$_{1-x}$Fe$_x$)OHFeSe~\cite{LiuFeng2018} have been tentatively
identified as Majorana bound
states (see Fig.~\ref{fig:Dispersions} c)). However, the
identification is still contraversial, and other groups
have questioned~\cite{ChenWen2018} whether the bound-states are 
conventional
Caroli-deGennes-Matricon~\cite{CaroliMatricon1964} states. 
Finally, a
robust zero bias peak, akin to a Majorana bound state was also
reported to occur at excess iron atoms of FeTe~\cite{YinPan2015} --
an effect possibly due to trapped fluxes ~\cite{JiangWang2018}. 
%From these arguments, built around 
%the topological bandstructure of Fig.~\ref{fig:Dispersions} a), we think
%that there is well-founded hope to realize 
These experimental observations provide the foundation 
our theoretical prediction of helical Majorana modes in
the vortex cores of FeSC Dirac semimetals. Moreover, a successful
experimental observation of 
helical Majorana modes in FeSC could be used
as independent experimental confirmation
of the topological paradigm proposed for FeSC.

Li(Fe$_{1-x}$Co$_x$)As, 
in which 3D bulk Dirac cones were observed in (S)ARPES at a
doping level of x = 0.09 \cite{ZhangShin2018}, is a strong candidate
for these Majorana modes. It exhibits  a $T_c(x =
0.09) \approx 9 K$ \cite{DaiDing2015} and, like all FeSC, is a strongly
type-II superconductor.  To get an insight of typical experimental
scales we compare to STM studies~\cite{HanaguriTakagi2012,ZhangHasan2019} of
vortices in the parent compound LiFeAs (here $T_c = 18K$ is larger but
comparable). Vortices are observable at $B \geq 0.1 T$ corresponding
to typical vortex spacing of $l_B \lesssim 80$ nm, while the core
radius is $\xi \approx 2.5$ nm.  Therefore, intervortex tunneling,
which would gap~\cite{LiuFranz2015} the zero modes is expected to be
weak. Furthermore, the large ratio  $\Delta/E_F \sim
0.5 \dots 1$ implies that the helical Majorana band should be well
separated in energy from conventional Caroli-deGennes-Matricon
states~\cite{CaroliMatricon1964}.

A pair
of helical Majorana modes displays universal thermal conductivity of
$\kappa_0 = \mathcal L T e^2/h$ where $\mathcal L = \pi^2 k_B^2/3e^2$
is the Lorenz number~\cite{PacholskiAdagideli2018}, and the
observation of this linear thermal conductivity is a key prediction of 
our theory. 
In the flux
phase, each of the $\Phi/\Phi_0$ vortices hosts 2 pairs of Majoranas,
so that the linear magnetic field dependence $\kappa_{\rm tot} = 2
\kappa_0 \Phi/\Phi_0$ of the total heat transport along the magnetic
field direction can be easily discriminated from the  phonon
background. A similar effect occurs in the specific heat $C = 2 c_0
\Phi/\Phi_0$ with $c_0 = \pi k_B^2 T/3 v_M$. Furthermore, STM
measurements are expected to detect a spatially localized signal in
the center of the vortex, with nearly constant energy dependence of
the tunneling density of states $\nu(E) \stackrel{E \rightarrow
0}{\simeq} 1/\pi v_M$.

\textit{Summary and Outlook.} In conclusion, we have demonstrated that
propagating Majorana modes are expected to develop in the
vortex cores of iron-based superconductors,  see Fig.~\ref{fig:Dispersions} e). These states are protected
by crystalline $C_4$ symmetry, but generic topological considerations,
Fig.~\ref{fig:Origin} and Eq.~\eqref{eq:IndexSemiclassical} suggest 
they will be robust against weak misalignments. A key signature of these
gapless excitations would be an 
dependence of various thermodynamic and transport
observables on the density of vortices and magnetic
field.

%We note that 
%analogous excitations at the edges of
%topological superconductors have been proposed
%as a way to implement \cite{HuZhao2018,LianZhang2018}
%braiding operations for topological quantum computing.  Another
We conclude with an 
interesting connection which derives from the close analogy
between superconducting and superfluid vortices 
and cosmic strings~\cite{VolovikBook}: 
line defects thought to be formed in the early universe in response
to spontaneous symmetry breaking of a grand unified field theory (GUT).  Defects
capable of trapping dispersive fermionic zero
modes~\cite{JackiwRossi1981} may occur in speculative SO(10) GUTs but
also in standard electroweak
theory~\cite{VilenkinShellard2000,Witten1985} and in either case the
interaction of cosmic strings with magnetic fields leads to a sizeable
baryogenesis. Helical Majorana modes in the vortex of FeSC may permit 
an experimental platform for testing these ideas. 

\textit{Note added.} Two
preprints~\cite{QinHu2019,QinHu2019b} appeared on the arXiv simultaneously to ours and  present consistent results on 1+1D
Majorana modes in vortices of FeSCs.

\section{Acknowledgements} We are grateful for discussions with
P.Y. Chang, H. Ding, V. Drouin-Touchette, Y. Komijani, P. Kotetes,
M. Scheurer and P. Volkov.  This work was supported by
the U.S. Department of Energy, Office of Basic
Energy Sciences, under Award DE- FG02-99ER45790 (Elio Koenig and Piers
Coleman) and by a QuantEmX travel grant (E. Koenig) from
the Institute for Complex Adaptive Matter and the Gordon and Betty Moore Foundation through Grant GBMF5305.

%\bibliography{MajoFeSC.bib}


%%%%%%%%%%%%%%%%%%%%%%%%%%%%%%%

\clearpage
\begin{widetext}

%%%%%%%%%% Prefix a "S" to all equations, figures, tables and reset the counter %%%%%%%%%%
\setcounter{equation}{0}
\setcounter{figure}{0}
\setcounter{table}{0}
\setcounter{page}{1}
\makeatletter
\renewcommand{\theequation}{S\arabic{equation}}
\renewcommand{\thefigure}{S\arabic{figure}}
\renewcommand{\thepage}{S\arabic{page}}
\renewcommand{\bibnumfmt}[1]{[S#1]}
%\renewcommand{\citenumfont}[1]{S#1} %EJK: I took out the S in front of citation for arXiv submission)
%%%%%%%%%% Prefix a "S" to all equations, figures, tables and reset the counter %%%%%%%%%%

\begin{center}
Supplementary materials for \\
\textbf{{Crystalline symmetry protected helical Majorana modes in the iron pnictides}}
\end{center}

These materials contain the mathematical details behind the main
text, including sections on the perturbative Majorana solution,
the index theorem, and a quasiclassical calculation which allows to compare to earlier works on $^3$He. 
All references in this supplement refer to the bibliography of the main text.

\section{Perturbative Majorana solution}
\label{sec:PertSol}

\subsection{Normal state effective  Hamiltonian}

We employ the low energy model introduced in
Refs.~\cite{XuZhang2016,ZhangShin2018}.  It is favorable to express
the normal state Hamiltonian using a basis of up and down spin orbital states
given by $(\vert \Phi_\uparrow\rangle , \vert \Phi_\downarrow\rangle
)$ where $\vert \Phi_\uparrow\rangle 
=(\ket{z, \uparrow}, \ket{(x {+} i y)z, \uparrow}, \ket{(x {-} i y)z ,
\uparrow}, \ket{xy, \uparrow})$ involves the up-spin $p_{z}$ state and
the three up spin $t_{2g}$ d-states,
while 
$\vert \Phi_ \downarrow\rangle  =(\ket{z,
\downarrow}, \ket{(x - i y)z , \downarrow}, \ket{(x + i y)z,
\downarrow}, \ket{xy, \downarrow})$ are their time-reversed partners. 
Note the different order of
$\ket{(x \pm i y)z}$ orbitals in the up and down spin states.  
The Hamiltonian then has the following
block structure 
\begin{equation}\label{eq:Supp:H0}
H(\v k) =  
\left (\begin{array}{c|c}
H_+ (\v k)& \Lambda \\ \hline
\Lambda^T & H_- (\v k)
\end{array} \right )
\equiv 
\left ( \begin{array}{cccc | cccc}
M_1 & {v} k_+ & - {v} k_- & 0 & \textbf{0} & \textbf{0} & \bar \lambda_3 & 0 \\ 
{v} k_- & M_2^+ & B^* & \bar \gamma k_+ & \textbf{0} & \textbf{0} & 0 &  - \bar \lambda_2 \\ 
- {v} k_+ & B &  M_2^- & \bar \gamma k_- & \bar \lambda_3 & 0 & 0 & 0 \\ 
0 & \bar \gamma k_- & \bar \gamma k_+ & M_3 & 0 & \bar \lambda_2 & 0 & 0	 \\ 
\hline
\textbf{0} & \textbf{0} & \bar \lambda_3 & 0 & M_1 & -{v} k_- &  {v} k_+ & 0 \\ 
 \textbf{0} & \textbf{0} & 0 & \bar \lambda_2 & -{v} k_+ & M_2^+ & B & \bar \gamma k_-\\ 
\bar \lambda_3 & 0 & 0 & 0 & {v} k_- & B^* &  M_2^- & \bar \gamma k_+  \\ 
0 & - \bar \lambda_2 & 0 & 0 & 0 & \bar \gamma k_+ & \bar \gamma k_- & M_3 
\end{array} \right) 
.
\end{equation}
We note that $H_-(\v k)= H_+^{T} (-\v k)$ is the
time-reversal of $H_+ (\v k)$. In this expression, the
transverse momenta $k_\pm = k_x \pm i k_y$, while $v$
is the transverse velocity associated with hybridization at the Dirac
point. In the notation of Ref.~\cite{ZhangShin2018} of the
main text, the various components of
the matrices are given by $v= \chi /\sqrt{2}$,
$\bar \gamma = \gamma\sin(k_z)/\sqrt{2}, M_{2}^\pm =
M_2 \pm \lambda_1, \bar \lambda_{2} = \lambda_2 \sqrt{2},\bar
\lambda_{3} = \lambda_3 \sin(k_z) \sqrt{2}, M_{n} = M_{n}^{(0)}
+ M^{(1)}_n(k_x^2 + k_y^2) +M^{(2)}_n 2(1 - \cos(k_z))$ ($n=1,2,3$), $B =
\beta k_+^2$.

In this basis, the Hamiltonian is
explicitly invariant under the tine-reversal operation 
$\theta = i\sigma_y K$ (where $K$ is the complex conjugation operator),
since the different ordering of up-spin and down-spin takes into
account the mapping of $\ket{(x \pm i y) z} \rightarrow \ket{(x \mp i
y) z}$. The Dirac node on the z-axis is seen to occur at the point where 
 $M_1 (\pm k_{z}^{*}) = M_2^+ (\pm k_{z}^{*})$ is apparent, since the
 spin-orbit submatrix (identified with boldface zeros), identically
 vanishes. 
%The quantity $v$ determins the 
%velocity of the electrons in directions transverse
% to the z-axis.  
The structure of the Hamiltonian
preserves the $z$ components of the total angular momentum $j_{z}$.
Thus  the states
$ \ket{z, \sigma}$ with  $j_z = \pm 1/2$ and the states 
$\ket{(x+i y)z,
\uparrow},\ket{(x-i y)z, \downarrow}$ with $j_{z} = \pm 3/2$ can not
mix.  
The block off-diagonal entries $\Lambda$ are the spin-flip
components of the spin-orbit coupling responsible for the topological
gap: thus the states $\vert z, \uparrow\rangle $ and $\vert x+iy ,
\downarrow \rangle $ with $j_{z}=1/2$ are mixed by the spin-orbit coupling
term $\bar{\lambda}_{3}$, while the states $\vert (x+iy)z, \uparrow\rangle $
and $\vert xy, \downarrow\rangle $ with $j_{z}=3/2$ are mixed
by the spin orbit coupling term $\bar{\lambda}_{2}$. 

In the main text, we introduce a Dirac Hamiltonian in Eq.~\eqref{eq:DiracHam}. We identify the parameters as $M_p \simeq M_1$, $M_d \simeq M_2^+$  up to corrections of higher order in $\bar \lambda_{2,3}$. Such corrections stem from the perturbative integration of orbitals which near the Dirac touching point are off-shell.

\subsection{Superconductivity}

In general, the inclusion of 
%spin-singlet 
superconductivity in the basis $(\Phi_{\v k}, i \sigma_y \Phi^*_{-\v k})$ ($\Phi$ being a multiorbital wave function) leads to
\begin{equation}
\mathcal H(\v k)  = \left (\begin{array}{cc}
H(\v k) -E_F & \Delta(\v k)  \\ 
\Delta^\dagger(\v k) & E_F - H(\v k) 
\end{array} \right ).
\end{equation}
%where we neglect the momentum dependence of the gaps $\Delta $. 
%Notice that the time-reversal invariance of $H (\v k)$ ensures that this term simply reverses under particle-hole transformations. 
Here, $E_F$ is the Fermi energy and, generally, $\Delta(\v k)$ is a matrix in orbital and spin space. The Pauli principle imposes $\Delta(\v k) = \sigma_y \Delta^T(-\v k) \sigma_y $, time reversal
symmetry, if present, implies then $\Delta(\v k) = \sigma_y \Delta^*(-\v k) \sigma_y = \Delta^\dagger(\v k)$. If
furthermore s-wave pairing is assumed, the gap function $\Delta (\v k) $ has the same symmetries as the Hamiltonian, i.e. a form analogous
to Eq.~\eqref{eq:Supp:H0}. For simplicity, we concentrate only on the
constant spin-singlet part, which is perfectly diagonal $\Delta =
\text{diag}(\Delta_1, \Delta_2^+, \Delta_2^-, \Delta_3)$. The gap functions introduced in the main text can be identified to leading order as $\Delta_p \simeq \Delta_1$ and $\Delta_d \simeq \Delta_2^+$.

In the presence of a vortex tube in the $(001)$ direction, $\Delta \rightarrow \Delta (r) e^{i \theta}$ and time reversal symmetry is broken. We use cylindrical coordinates $(x,y,z) = (r \cos(\theta), r \sin(\theta), z)$. The particle-hole symmetry of the superconducting Hamiltonian $C \mathcal H^* C = - \mathcal H$ with $C = \sigma_y \tau_y$ persists even in the presence of the vortex. In view of rotational symmetry, it is possible to assign the quantum numbers $k_z$ (momentum in $z$-direction), $l$ (angular momentum in $x-y$ plane) and $n$ (radial quantum number) in the bulk of the system. Particle hole symmetry implies the appearance of pairs of eigenstates $\Psi_{k_z, n, l}(\v x)$ and $\sigma_y \tau_y \Psi^*_{k_z, n, l}(\v x) \propto \sigma_y \tau_y \Psi_{-k_z, n, -l}(\v x)$ with opposite eigenenergy $E_{k_z, n, l} = -E_{-k_z, n, -l}$. Furthermore, there is an inversion symmetry $z \rightarrow -z, \theta \rightarrow \theta + \pi$ which is represented by $P = \diag{-\mathbf 1_\sigma,\mathbf 1_\sigma,\mathbf 1_\sigma,\mathbf 1_\sigma;\mathbf 1_\sigma,-\mathbf 1_\sigma,-\mathbf 1_\sigma,-\mathbf 1_\sigma}$ and implies degeneracy of $E_{k_z, n, l}$ and $E_{-k_z, n, l}$.

\subsection{Majorana solutions}

We now switch to the basis $(\Phi_\uparrow, \Phi_\downarrow^*, \Phi_\downarrow, -\Phi_\uparrow^*)$, this corresponds to 
an additional rotation in Nambu space swapping second and third blocks. We obtain

\begin{equation}
\mathcal H = \left (\begin{array}{cc|cc}
H_+ - E_F  & \Delta & \Lambda & {0} \\ 
\Delta^* & E_F - H_+ & {0} & - \Lambda \\ 
\hline
\Lambda^T & {0} & H_- - E_F & \Delta \\ 
{0} & -\Lambda^T & \Delta^* & E_F - H_-
\end{array} \right ). \label{eq:switchedNambu}
\end{equation}

The topologically most interesting features of the bulk spectrum are
given by the crossing $M_1(k^{*}_z) = M_{2}^+(k^{*}_z)$ (Dirac point) and the
anticrossing $M_1(k'_z) = M_{2}^-(k'_z)$ (topological gap). Since
$M_{3}$ is finite at these points, we can drop the $d_{xy}$
states, reducing the dimension of the block-diagonal matrices to
three. 
In this section we absorb the  group velocity $v$ of the Dirac
cone in the x,y directions, into redefined length
scales, replacing $(x,y)/v\rightarrow (x,y)$ 
such that $v k_\pm = -i (\partial_x \pm i
\partial_y)$. We further set $\bar \beta = \beta/ v^2$, $\bar
M_n = M_n^{(0)} + M_n^{(2)} 2(1- \cos(k_z)) -
({M_n^{(1)}/v^2}) (\partial_x^2 +\partial_y^2) - E_F$, (n = 1,2,3)
so that
\begin{align}
H_+ &= \left (\begin{array}{ccc}
\bar M_1 & p_+ & -p_- \\ 
p_- &\bar M_2^+ & \bar \beta p_-^2 \\ 
- p_+ & \bar \beta p_+^2 &\bar M_2^- 
\end{array} \right ) , & 
H_- &= \left (\begin{array}{ccc}
\bar M_1 & -p_- & p_+ \\ 
-p_+ &\bar M_2^+ & \bar \beta p_+^2 \\ 
p_- & \bar \beta p_-^2 &\bar M_2^- 
\end{array} \right )\\
\Lambda &= \left (\begin{array}{ccc}
0 & 0 & \bar \lambda_3 \\ 
0 & 0 & 0 \\ 
\bar \lambda_3 & 0 & 0
\end{array} \right ), & 
\Delta &= \left (\begin{array}{ccc}
\Delta_1(r) e^{i \theta} & 0 &0 \\ 
0 & \Delta_2^+(r) e^{i \theta} & 0 \\ 
0 & 0 & \Delta_2^-(r) e^{i \theta}
\end{array} \right ),
\end{align}
%where the omission of the $d_{xy}$ states has made the blocks three dimensional. 
The Zeeman field adds $\delta \mathcal H = g \mu_B B/2 \text{diag}(\mathbf 1, \mathbf 1, - \mathbf 1, -\mathbf 1)$ to Eq.~\eqref{eq:switchedNambu}.

The emergent rotational invariance of the effective Hamiltonian allows
us to expand the wave functions at a given $k_z$ in angular momenta $l$
\begin{equation}
\Psi_\pm (x,y) = \sum_l e^{i l \theta }U_\pm(\theta) \Psi^{(l)}_\pm(r)
\end{equation}
with
\begin{eqnarray}
U_+(\theta) &=& \diag{e^{ - i \pi/4}, e^{-i \theta + i \pi/4}, e^{i \theta + i \pi/4}, e^{-i\theta-i \pi/4}, e^{-2i \theta+ i \pi/4}, e^{i \pi/4} },\\
U_-(\theta) &=& \diag{e^{i \theta + i \pi/4}, e^{2i \theta - i \pi/4}, e^{ - i \pi/4}, e^{ i \pi/4}, e^{i \theta - i \pi/4}, e^{-i \theta - i \pi/4}} .
\end{eqnarray}
The relative factors of $e^{i \theta}$ in various matrix elements reflect that different orbitals transform differently under rotations. The choice of phases of $\pi/4$ is pure convenience. 

Using this transformation we obtain in the $l$th sector
\begin{subequations}
\begin{equation}
\mathcal H^{(l)} = \left (\begin{array}{cc|cc}
H_+^{(l)} & \bar \Delta & \Lambda & {0} \\ 
\bar \Delta & - H_+^{(l{-1})} & {0} & - \Lambda \\ 
\hline
\Lambda & {0} & H_-^{(l{+1})} &\bar \Delta \\ 
{0} & -\Lambda & \bar \Delta & - H_-^{(l)}
\end{array} \right ),
\end{equation}
with $\bar \Delta = \diag{\Delta_1,\Delta_2^+,\Delta_2^-}$ and
\begin{align}
H_+^{(l)} &= \left (\begin{array}{ccc}
\bar M_1^{(l)} &  D_r^{1 - l} & - D_{r}^{1+l} \\ 
- D^{l}_r & \bar M_2^{(l-1),+} & - \bar \beta D_r^{l}D_r^{l+1} \\ 
 D_r^{-l} & - \bar \beta D_r^{-l} D_r^{1-l}& \bar M_2^{(l+1),-}
\end{array} \right ) , & 
H_-^{(l)} &=\left (\begin{array}{ccc}
\bar M_1^{(l)} &  D_{r}^{1+l} & -  D_r^{1 - l} \\ 
- D^{-l}_r & \bar M_2^{(l+1),+} &  - \bar \beta D_r^{-l}  D_r^{1-l}\\ 
 D_r^{l} & - \bar \beta D_r^{l} D_r^{l+1}& \bar M_2^{(l-1),-}
\end{array} \right ). 
\end{align}
\label{eq:HamiltonianCylCoord}
\end{subequations}
Here we have introduced $D_r^{k} =\partial_r +k/r$ and $\bar M_i^{(k)} =
\bar M_i \vert_{M^{(1)}_i = 0} - M_i^{(1)}[D_r^{1-k}D_r^k + D_r^{1+k}
D_r^{-k}]/[2v^2]$. The shift by $l = 1$ between particle and
hole sectors is a consequence of chosing a vortex with winding $+1$.
We remind ourselves that in the inner product in cylindrical coordinates is $\braket{\Psi \vert \Phi} = 2\pi \sum_l \int dr r \Psi^*_l (r) \Phi_l(r)$. This is the reason why Eq.~\eqref{eq:HamiltonianCylCoord} appears non-Hermitian (it is self-adjoint but with respect to the above inner product). 
To make hermiticity apparent, one may define $\Psi^{(l)} (r) = \tilde \Psi(r)^{(l)} /\sqrt{2\pi r}$ where $\tilde \Psi$ lives in a spinorial Hilbert space with usual $L^2$ norm. Clearly $\mathcal H^{(l)} \Psi^{(l)} = E \Psi^{(l)}$ implies $\tilde {\mathcal H}^{(l)} \tilde \Psi^{(l)} = E \tilde \Psi^{(l)}$ and the hermitian Hamiltonian $\tilde{\mathcal H}^{(l)}$ takes the form
\begin{subequations}
\begin{equation}
\tilde{\mathcal H}^{(l)} = \sqrt{r}{\mathcal H}^{(l)} \frac{1}{\sqrt{r}} = \left (\begin{array}{cc|cc}
\tilde H_+^{(l)} & \bar \Delta & \Lambda & {0} \\ 
\bar \Delta & - \tilde H_+^{(l{-1})} & {0} & - \Lambda \\ 
\hline
\Lambda & {0} & \tilde H_-^{(l{+1})} &\bar \Delta \\ 
{0} & -\Lambda & \bar \Delta & - \tilde H_-^{(l)}
\end{array} \right ),
\end{equation}
with 
$\tilde M_i^{(k)} =
\bar M_i \vert_{M^{(1)}_i = 0} - M_i^{(1)}[D_r^{1/2-k}D_r^{k-1/2} + D_r^{1/2+k}
D_r^{-k-1/2}]/[2v^2]$ and
\begin{align}
\tilde H_+^{(l)} &= \left (\begin{array}{ccc}
\tilde M_1^{(l)} &  D_r^{1/2 - l} &- D_{r}^{1/2+l} \\ 
- D^{l-1/2}_r & \tilde M_2^{(l-1),+} & - \bar \beta D_r^{l-1/2}D_r^{l+1/2} \\ 
D_r^{-l-1/2} & - \bar \beta D_r^{-l-1/2} D_r^{1/2-l}& \tilde M_2^{(l+1),-}
\end{array} \right ), & 
\tilde H_-^{(l)} &=\left (\begin{array}{ccc}
\tilde M_1^{(l)} & D_{r}^{1/2+l} & -  D_r^{1/2 - l} \\ 
- D^{-l-1/2}_r & \tilde M_2^{(l+1),+} &  - \bar \beta D_r^{-l-1/2}  D_r^{1/2-l}\\ 
 D_r^{l-1/2} & - \bar \beta D_r^{l-1/2} D_r^{l+1/2}& \tilde M_2^{(l-1),-}
\end{array} \right ) 
\end{align}
\label{eq:HamiltonianCylCoordHerm}
\end{subequations}

To make further progress we return to the more standard representation ${\mathcal H}$ and now concentrate on the two most interesting situations when the chemical potential is near the Dirac point or near the topological anticrossing.

\subsubsection{Case 1: Fermi energy near topological anticrossing}

	We begin the discussion by concentrating on the regime where the chemical potential is in the vicinity of the topological anticrossing. To find a perturbative solution, we first concentrate on Eq.~\eqref{eq:HamiltonianCylCoord} near $\bar M_1\vert_{M_1^{(1)} = 0}(k_z) = \bar M_2^-\vert_{M_2^{(1)} = 0}(k_z)$ in the approximation of linearized momenta $\beta = M_1^{(1)} = M_2^{(1)} = 0$, setting $\bar \lambda_3= 0$ and projected onto the relevant bands, i.e. $(\ket{z, \uparrow},\ket{(x-iy)z, \uparrow},\ket{z, \downarrow},\ket{(x+iy)z, \downarrow}$. We furthermore introduce ``center of mass'' $\Delta(r) = \frac{\Delta_1(r) + \Delta_2^-(r)}{2}$ and relative pairing gaps $\delta \Delta(r) = \Delta_1(r) - \Delta_2^-(r)$. The zeroth order Hamiltonian is a direct sum of $+$ and $-$ sectors	
\begin{align} \label{eq:TIAvCrossing}
\mathcal H^{(l)}_{+,0} &= \left (\begin{array}{cccccc}
- \mu & 0 & -D_r^{1+l} & \Delta(r) & 0 & 0 \\ 
0 & 0 & 0 & 0 & 0 & 0 \\ 
D_r^{-l} & 0 & -\mu & 0 & 0 & \Delta(r) \\ 
\Delta(r) & 0 & 0 & \mu & 0 & D_r^{l} \\ 
0 & 0 & 0 & 0 & 0 & 0 \\ 
0 & 0 & \Delta(r) & -D_r^{1-l} & 0 & \mu
\end{array}  \right ), 
& \mathcal H^{(l)}_{-,0} &= \left (\begin{array}{cccccc}
- \mu & 0 & -D_r^{l} & \Delta(r) & 0 & 0 \\ 
0 & 0 & 0 & 0 & 0 & 0 \\ 
D_r^{1+l} & 0 & -\mu & 0 & 0 & \Delta(r) \\ 
\Delta(r) & 0 & 0 & \mu & 0 & D_r^{1-l} \\ 
0 & 0 & 0 & 0 & 0 & 0 \\ 
0 & 0 & \Delta(r) & -D_r^{l} & 0 & \mu
\end{array}  \right ), 
\end{align}
where $\mu = E_F - 
\left( 
{\bar M_1\vert_{M_1^{(1)} = 0}+ \bar M_2^-\vert_{M_2^{(1)} = 0}}\right)/2
$ .

We readily find that a chiral symmetry

\begin{equation}
\tau_y\left (\begin{array}{ccc}
0 & 0 & -i \\ 
0 & 0 & 0 \\ 
i & 0 & 0
\end{array} \right ) \mathcal H_{+/-,0} \tau_y  \left (\begin{array}{ccc}
0 & 0 & -i \\ 
0 & 0 & 0 \\ 
i & 0 & 0
\end{array} \right ) =- \mathcal H_{+/-,0}
\end{equation}
exists if and only if $l = 0$ in both up and down spin sectors. This chiral symmetry is the necessary ingredient for the determination of the zero energy Majorana mode in Eq.~\eqref{eq:TIAvCrossing}. The zeroth order wave functions are thus 
	
	\begin{align}
		\Psi_+^{(l = 0)} (r) &= \mathcal N e^{- \int_0^r \Delta(r') dr'} \left (\begin{array}{c}
		J_0(\mu r) \\ 
		0\\ 
		J_1(\mu r)  \\ 
		J_1(\mu r)  \\ 
		0 \\ 
		-J_0(\mu r) 
		\end{array} \right ), & 
		\Psi_-^{(l = 0)} (r) &= \mathcal N e^{- \int_0^r \Delta(r') dr'} \left (\begin{array}{c}
		J_1(\mu r)  \\ 
		0 \\ 
		-J_0(\mu r)  \\ 
		-J_0(\mu r) \\ 
		0 \\ 
		-J_1(\mu r)
		\end{array} \right )
	\end{align}
	
\subsubsection{Case 2: Fermi energy near the Dirac point}

We now switch to the regime where the chemical potential is in the
vicinity of the topological Dirac semimetal. To find a perturbative
solution, we now concentrate on Eq.~\eqref{eq:HamiltonianCylCoord}
near $\bar M_1\vert_{M_1^{(1)} = 0}(k_z) = \bar M_2^+\vert_{M_2^{(1)}
= 0}(k_z)$, again in the approximation of linearized momenta $\beta =
M_1^{(1)} = M_2^{(1)} = 0$, setting $\bar \lambda_3= 0$ and projected
onto the relevant bands, which in this case are $(\ket{z,
\uparrow},\ket{(x+iy)z, \uparrow},\ket{z, \downarrow},\ket{(x-iy)z,
\downarrow}$. We use slightly different notation to the previous
section $\Delta(r) = \frac{\Delta_1(r) + \Delta_2^+(r)}{2}$, $\delta
\Delta(r) = \Delta_1(r) - \Delta_2^+(r)$, $\mu = E_F -
\left( 
{\bar M_1\vert_{M_1^{(1)} = 0}+ \bar M_2^+\vert_{M_2^{(1)} = 0}}\right)/{2}$. We remark that, at $k_z^*$ this definition of the chemical potential is the same as the one employed in the main text (in view of the tilt, the two definitions are not exactly equivalent, but differences in the effective helical Majorana Hamiltonian appear only at second order in perturbation theory, i.e. they are beyond the level of accuracy of this calculation). The zeroth order Hamiltonian is again a direct sum of $\uparrow$ and $\downarrow$ sectors	
\begin{align} \label{eq:TDSCrossing}
\mathcal H^{(l)}_{+,0} &= \left (\begin{array}{cccccc}
- \mu & D_r^{1-l} & 0 & \Delta(r) & 0 & 0 \\ 
-D_r^{l}  &  -\mu  & 0 & 0 & \Delta(r) & 0 \\ 
0 & 0 &0 & 0 & 0 & 0 \\ 
\Delta(r) & 0 & 0 & \mu & -D_r^{2-l} & 0 \\ 
0 & \Delta(r)  & 0 &  D_r^{l-1}  & \mu & 0 \\ 
0 & 0 & 0 & 0 & 0 & 0
\end{array}  \right ), 
& \mathcal H^{(l)}_{-,0} &= \left (\begin{array}{cccccc}
- \mu & D_r^{l+2} & 0 & \Delta(r) & 0 & 0 \\ 
-D_r^{-l-1}  &  -\mu & 0 &0 &  \Delta(r) & 0 \\ 
0 & 0 & 0 & 0 & 0 & 0 \\ 
\Delta(r) & 0 & 0 & \mu & -D_r^{1+l} & 0 \\ 
0 & \Delta(r) & 0 & D_r^{-l} & \mu & 0 \\ 
0 & 0 &  & 0  & 0 & 0 
\end{array}  \right ), 
\end{align}

In this energy and momentum regime, we find a chiral symmetry

\begin{equation}
\tau_y\left (\begin{array}{ccc}
0 & -i & 0\\ 
i & 0 & 0 \\ 
0 & 0 & 0
\end{array} \right ) \mathcal H_{+/-,0} \tau_y  \left (\begin{array}{ccc}
0 & -i & 0 \\ 
i & 0 & 0 \\ 
0 & 0 & 0
\end{array} \right ) =- \mathcal H_{+/-,0}
\end{equation}
which exists if and only if $l = 1$ ($l = -1$) in the + (-) helicity sectors. Keeping in mind that the chiral symmetry is necessary ingredient for the zero energy solution in Eq.~\eqref{eq:TDSCrossing}, the Majorana wave functions are thus in sectors of different angular momentum and therefore
	
	\begin{align}
		\Psi_+^{(l = 1)} (r) &= \mathcal N e^{- \int_0^r \Delta(r') dr'} \left (\begin{array}{c}
		-J_1(\mu r) \\ 
		J_0(\mu r) \\ 
		0 \\ 
		J_0(\mu r) \\ 
		J_1(\mu r) \\ 
		0
		\end{array} \right ) , &
		\Psi_-^{(l = -1)} (r) &= \mathcal N e^{- \int_0^r \Delta(r') dr'} \left (\begin{array}{c}
		J_0(\mu r) \\ 
		J_1(\mu r) \\ 
		0 \\ 
		J_1(\mu r) \\ 
		-J_0(\mu r) \\ 
		0
		\end{array} \right ) 
	\end{align}
We have also explicitly checked that the chiral symmetry is only 
present for the case of a vortex with odd winding number. The chiral
symmetry is absent for even winding, which  prevents helical Majorana
modes in these cases. 

\subsection{Approximate dispersion relations}

We now use the previously derived low-energy solutions to determine the effective Hamiltonian of Majorana vortex states. We use $\Delta_i(r) = \Delta_i^\infty \tanh(r/\xi)$ and $ (\Delta_1^\infty + \Delta_2^\infty) \xi\equiv 2\Delta^\infty \xi = 2$ with $i = (1,2,3) = (1,2_+,2_-)$ and $\Delta_2^+ = \Delta_2^-$ in the following - but the qualitative aspects are expected to be insensitive to this precise choice.
We further define the following integrals
\begin{align}
 I_+(x) &= \int_0^\infty dr \frac{r}{[\cosh(r/x)]^2} (J_0(r)^2+J_1(r)^2) \simeq \frac{\left| x\right|  \left(2-\pi  \left| x\right|  \log \left(\cosh \left(\frac{2}{\pi  x}\right)\right)\right)}{\pi }, \\
 I_-(x) &= \int_0^\infty dr \frac{r}{[\cosh(r/x)]^2} (J_0(r)^2-J_1(r)^2) \simeq \begin{cases} x^2 \ln(2), & x\ll 1 \\ \frac{1}{4x}, & x \gg 1,\end{cases} \\
 I_0(x) &= \int_0^\infty dr \frac{r \sinh(r/x)}{[\cosh(r/x)]^3
} 2J_0(r) J_1(r) = x I_-(x).
\end{align}

The expectation value of the full Hamiltonian with respect to the wave function of Majorana solution leads to the first perturbative low energy Hamiltonian (we here present only the case of linearized dispersion). For case 1, in the basis $\lbrace \Psi_+^{(l= 0)},\Psi_-^{(l= 0)}\rbrace $ we obtain
\begin{equation}
\mathcal H \simeq \left (\begin{array}{cc}
 (M_1 - M_2^-) \frac{I_-(\frac{\mu}{\Delta^\infty})}{2 I_+(\frac{\mu}{\Delta^\infty})} + (\Delta_1^\infty - \Delta_2^\infty) \frac{I_0(\frac{\mu}{\Delta^\infty})}{2 I_+(\frac{\mu}{\Delta^\infty})}  +\frac{g \mu_B B}{2} & -\bar \lambda_3 \frac{I_-(\frac{\mu}{\Delta^\infty})}{ I_+(\frac{\mu}{\Delta^\infty})}  \\ 
-\bar \lambda_3 \frac{I_-(\frac{\mu}{\Delta^\infty})}{ I_+(\frac{\mu}{\Delta^\infty})}  & - (M_1 - M_2^-) \frac{I_-(\frac{\mu}{\Delta^\infty})}{2 I_+(\frac{\mu}{\Delta^\infty})} - (\Delta_1^\infty - \Delta_2^\infty) \frac{I_0(\frac{\mu}{\Delta^\infty})}{2 I_+(\frac{\mu}{\Delta^\infty})} - \frac{g \mu_B B}{2}
\end{array} \right ).
\end{equation}

In contrast, for case 2, i.e. a Fermi energy near the Dirac crossing, we find in the basis of $\lbrace \Psi_+^{(l= +1)},\Psi_-^{(l= -1)}\rbrace $
\begin{equation}
\mathcal H \simeq \left (\begin{array}{cc}
- (M_1 - M_2^+) \frac{I_-(\frac{\mu}{\Delta^\infty})}{2 I_+(\frac{\mu}{\Delta^\infty})} - (\Delta_1^\infty - \Delta_2^\infty) \frac{I_0(\frac{\mu}{\Delta^\infty})}{2 I_+(\frac{\mu}{\Delta^\infty})} +\frac{g \mu_B B}{2}  & 0  \\ 
0  &  (M_1 - M_2^+) \frac{I_-(\frac{\mu}{\Delta^\infty})}{2 I_+(\frac{\mu}{\Delta^\infty})} + (\Delta_1^\infty - \Delta_2^\infty) \frac{I_0(\frac{\mu}{\Delta^\infty})}{2 I_+(\frac{\mu}{\Delta^\infty})} - \frac{g \mu_B B}{2}
\end{array} \right ).
\end{equation}
We thus observe that near the topological anticrossing, Majorana modes of $l = 0$ mutually gap out, Fig. \ref{fig:Dispersions} b) of the main text, while at the Dirac point, $C_4$ symmetry protects the appearance of helical Majorana modes Fig. \ref{fig:Dispersions} d) of the main text. This section also concludes the derivation of the velocity $v_M \stackrel{\vert \mu \vert \gg \Delta^{\infty}}{\sim} (\Delta^\infty/\mu)^2 \partial_{k_z}\vert_{k_z^*} [M_2 - M_1^+]$ of the helical Majorana modes. An analogous result with the same factor $(\Delta^\infty/\mu)^2$ and obtained by different means for the case of a the o-vortex in $^3$He-B was presented in Eq.~(7.2) of Ref.~\cite{MisirpashaevVolovik1995}. We also highlight that in the limit $\vert \mu\vert \ll \Delta^{\infty}$ the velocity is $v_M \simeq \partial_{k_z}\vert_{k_z^*} [M_2 - M_1^+]/2$.

\section{Index Theorem}
\label{sec:Index}

In this section we summarize the semiclassical evaluation of the index, Eq.~\eqref{eq:IndexMainText} of the main text, which ensures the appearance of propagating Majorana fermions. The definition and the idea of a semiclassical evaluation of the index follows Ref.~\cite{Volovik1989} for superfluid $^3$He. However the connection to the Berry curvature monopoles and spin Hall conductance was not drawn in that context.

In contrast to all other parts of this work, $z$ here denotes the direction of the vortex line and is in general not the same as the \textit{c} axis of the crystal.

\subsection{Non-degenerate Fermi surface - semiclassical expansion}

Following the explanations of the main text, we consider a Bogoliubov-de Gennes Hamiltonian of each band separately, i.e. $N(k_z) = \sum_\xi \sum_\pm \mp \tilde N_{\xi, \pm}(k_z)/2$. We have assumed absent interband pairing and we explicitly checked that interband contributions, which are induced by the spatial dependence of $\Delta(\v r)$, vanish from $N(k_z)$ at the leading order in gradient expansion. For the sake of a more transparent notation, we suppress the $\xi$ and $\pm$ indices and treat each non-degenerate system separately
\begin{equation}
\mathcal{H} = {\v d} \cdot \boldsymbol \tau.
\end{equation}
The object ${\v d}$ is a three vector of which each component is an operator in real/momentum space and $\boldsymbol \tau$ representing Pauli matrices in Nambu space. %We have at this point omitted band (helicity) indices $\xi$ ($\pm$). 
We deform the contour of integration in Eq.~\eqref{eq:IndexMainText} as follows
\begin{equation}
\tilde N(k_z) = \text{Im} \int_{- \infty}^0 \frac{d\omega}{\pi} e^{\omega 0^+}\tr [(\omega - i 0 - \mathcal H)^{-1}] = \int_{- i\infty}^{i\infty} \frac{d z}{2\pi i} e^{z 0^+} \tr[\underbrace{(z - \mathcal H)^{-1}}_{=\mathcal G(z)}].
\end{equation}
The ``$\tr$'' operation denotes trace in the entire Hilbert space at given $k_z$, and can be visualized as the trace in Nambu space and momentum space transversal to $z$.
%As we readily show, such a Hamiltonian may be obtained by projecting Eq.~\eqref{eq:DiracHam} or any other Hamiltonian with non-degenerate bands onto the conduction band.
We now use ``$\circ$'' to denote operator convolution (e.g. in momentum space) and expand $\mathcal G$ to leading order in the quantum commutators
\begin{eqnarray}
\mathcal G(z) & = & [z - \v d \cdot \boldsymbol \tau]^{-1}  = \frac{1}{2} \lbrace (z + \v d \cdot \boldsymbol \tau) \KC [z^2 - \v d^2 - i \epsilon_{abc} \tau_a [d_b \KC d_c]/2]^{-1}\rbrace  \notag \\
&\simeq & \frac{1}{2} \lbrace (z + \v d \cdot \boldsymbol \tau) \KC [z^2 - \v d^2]^{-1} +\frac{ i}{2} \epsilon_{abc} \tau_a  [z^2 - \v d^2]^{-1} \circ [d_b \KC d_c] \circ [z^2 - \v d^2]^{-1} \rbrace.
\end{eqnarray}

We thus obtain
\begin{eqnarray}
\tilde N(k_z) &=& \int_{-\infty}^\infty \frac{d \epsilon}{2\pi} e^{i \epsilon 0^+} \sum_{\v p} \lbrace \frac{-i}{\epsilon^2 + \v d^2} \rbrace_{\v p, \v p} \notag \\
&+& \int_{-\infty}^\infty \frac{d \epsilon}{2\pi} e^{i \epsilon 0^+} \sum_{\v p} \frac{i \epsilon_{abc}}{2} \lbrace  d_{c} \KC [\epsilon^2 + {\v d}^2]^{-1} \circ [ d_{a} \KC  d_{b}] \circ [\epsilon^2 + {\v d}^2]^{-1}  \rbrace_{\v p, \v p}
\end{eqnarray}

Here, momentum space has been used to visualize the meaning of the trace operation. We anticipate that in the semiclassical approximation, the first line yields the same, $k_z$ independent result in both helicity sectors and thus drops out of the difference $N_+ - N_- $. We will disregard it henceforth. The convergence factor in the second line can be dropped as the $\epsilon$ integral converges.

\subsection{Standard Moyal product and a simplified case}
Before turning to the generlized Wigner transform introduced in the main text, we make use of the concepts of the standard Wigner transformation and Moyal product, see e.g. A. Kamenev, \textit{Field Theory of Non-Equilibrium Systems}, Cambridge University Press (2011). For arbitrary operators $\hat A, \hat B$ this implies
\begin{align}
A(\v R, \v P) &= \int \frac{d^2 k}{(2\pi)^2} A(\v P + \frac{\v k}{2},\v P - \frac{\v k}{2}) e^{i \v k \cdot \v R},  \\
[A \circ B](\v R, \v P) &= A(\v R, \v P) e^{\frac{i}{2}\left ( \overleftarrow{\nabla}_X \overrightarrow{\nabla}_P- \overleftarrow{\nabla}_P \overrightarrow{\nabla}_X \right )} B(\v R, \v P),
\end{align}
where $\circ$ denotes subsequent application of operators. Derivatives in real and momentum space $\nabla_X$ and $\nabla_P$ acting to the left (right) are denoted by arrows $\leftarrow$ ($\rightarrow$) in the superscript. Leading order expansion in gradients leads to

\begin{equation}
\tilde N(k_z) = - \frac{1}{2} \int \frac{d^2P d^2X}{(2\pi)^2} \epsilon_{abc} \frac{ d_{a}(\v R, \v P) }{d(\v R, \v P)}\vec \nabla_X  \frac{d_{b}(\v R, \v P) }{d(\v R, \v P)}\cdot \vec \nabla_P  \frac{d_{c}(\v R, \v P)}{d(\v R, \v P)}. \label{eq:SemiclassicalIndex}
\end{equation}

We first consider the simplified case where we linearize a generic isotropic vortex in an orbital independent order parameter field. We consider a Hamiltonian of the form
\begin{equation}
\mathcal H = \left (\begin{array}{cc}
H_{\v p} - \mu & \Delta_\infty \left (\frac{x}{\xi} - i \frac{y}{\xi} \right) \\ 
\Delta_\infty \left (\frac{x}{\xi} + i \frac{y}{\xi}\right) & \mu - H_{\v p} 
\end{array} \right ). \label{eq:Index:LinearizedModel}
\end{equation}
Projected onto the band with states $\ket{u_{\v p}}$, we obtain $\mathcal H = {\v d} \cdot \boldsymbol \tau$ with
\begin{equation}
 {\v d} = (\Delta_\infty [i\partial_{p_x} + \mathcal A_x]/\xi,\Delta_\infty [i\partial_{p_y} + \mathcal A_y]/\xi, \epsilon_{\v p} - \mu). 
\end{equation} 
and $\mathcal A_{x,y} = i\braket{u_{\v p} \vert \partial_{p_{x,y}} u_{\v p}}$ denotes the Berry connection. 
With the above mentioned Wigner transform we obtain $\v d(\v R, \v P) = (\Delta_\infty (X + \mathcal A_x)/\xi, \Delta_\infty (Y + \mathcal A_y)/\xi, \epsilon_{\v p} - \mu)$. We use that the contribution of $c = z$ to Eq.~\eqref{eq:SemiclassicalIndex}
\begin{equation}
\int \frac{d^2P d^2X}{(2\pi)^2} \epsilon_{abz} \frac{d_a(\v R, \v P)}{ d(\v R, \v P)} \nabla_X  \frac{d_b(\v R, \v P)}{ d(\v R, \v P)} \cdot \nabla_P  \frac{d_z(\v R, \v P)}{ d(\v R, \v P)} = \int \frac{d^2P d^2X}{(2\pi)^2} \epsilon_{abz}  \frac{d_a(\v R, \v P)}{ d(\v R, \v P)^3} v_b(\v P) \frac{\Delta_\infty}{\xi} = 0
\end{equation}
where at the last equality sign we took the $\v R$ integral first and shifted $\v R \rightarrow (\v R - \boldsymbol{\mathcal A})/\xi$. 
Then, we obtain
\begin{equation}
\tilde N(k_z) =- \frac{\Delta_\infty^2}{2 \xi^2} \int \frac{d^2P d^2X}{(2\pi)^2} \frac{d_3(\v R, \v P)}{d^3(\v R, \v P)} \Omega_z. \label{eq:Index:IndexSimpleModel}
\end{equation}

\subsection{Gauge invariant Wigner transform and generic case}
For a more generic coordinate dependence of the order parameter it is advantageous to define a Wigner transform which respects the gauge invariance $\ket{u_{\v p}} \sim e^{i \phi_{\v p}} \ket{u_{\v p}}$ of eigenstates. Starting from the projection $A(\v p, \v p') = \braket{u_{\v p} \vert \hat A(\v p, \v p') \vert u_{\v p'}}$ of an orbital matrix $\hat A(\v p, \v p')$ onto a single, given band we define
\begin{equation}
A(\v R, \v P) = \int \frac{d^2 k}{(2\pi)^2} A(\v P + \v k/2, \v P- \v k/2) e^{i \v k (\v R - \boldsymbol{\mathcal A} (\v P))}.
\end{equation}
This Wigner transform is gauge invariant to zeroth and first order in gradient expansion. As a consequence, the Moyal product takes the form
\begin{align}
[A \circ B](\v R, \v P) &\simeq A(\v R, \v P) B(\v R, \v P) + \frac{i}{2} \left (\vec \nabla_X A \cdot \vec \nabla_P B - \vec \nabla_P A \cdot \vec \nabla_X B\right ) + \frac{i}{2} \Omega_z \hat e_z \cdot \left (\vec \nabla_X A \times \vec \nabla_X B\right ). \label{eq:MoyalGeneralizedWT}
\end{align}
Since we here concentrate on a 2D problem for each $k_z$ separately, in the anomalous last term only $\Omega_z$ enters. Using this definition of the Wigner transform and a Hamiltonian of the form
\begin{equation}
\mathcal H = \left (\begin{array}{cc}
H_{\v p} - \mu & \Delta(\v r)  \\ 
\Delta^*(\v r) & \mu - H_{\v p} 
\end{array} \right ). 
\end{equation}
we have
\begin{equation}
\v d(\v R, \v P) = (\text{Re}\Delta(\v R), - \text{Im}\Delta(\v R), \epsilon({\v P}) - \mu).
\end{equation}
Then we find
\begin{eqnarray}
\tilde N( k_z) = - \frac{1}{4} \int \frac{d^2P d^2X}{(2\pi)^2} \epsilon_{abc} \lbrace 2 \hat d_a \vec \nabla_X \hat d_b \cdot \vec \nabla_P \hat d_c + \hat d_a (\vec \nabla_X \hat d_b \times \vec \nabla_X \hat d_c) \cdot \vec \Omega \rbrace.
\end{eqnarray}
Unit vectors $\hat d = \v d/d$ are denoted with a hat. Straightforward inspection of this equation for the simplified model Eq.~\eqref{eq:Index:LinearizedModel} reproduces Eq.~\eqref{eq:Index:IndexSimpleModel} and demonstrates the validity of this generalized Wigner transformation. Furthermore, since the first term is independent of helicity, it drops out of the final index $[N_+(k_z) - N_-(k_z)]/2$.
We return to a more generic gap function and disregard interband pairing, so that $\Delta(\v R) = [(X+iY)/R]^{\nu_v} \vert \Delta(\v R) \vert$. In this case
\begin{equation}
\vec \nabla_X d_x \times \vec \nabla_X d_y = -\frac{\nu_v}{2R} \partial_R \vert \Delta (R) \vert^2 \hat e_z.
\end{equation} 
\begin{subequations}
\begin{eqnarray}
\tilde N(k_z) &=& \frac{\nu_v}{4} \int \frac{d^2P d^2X}{(2\pi)^2} \epsilon_{abc} \hat d_a (\vec \nabla_X \hat d_b \times \vec \nabla_X \hat d_c) \cdot \vec \Omega  \\
%&=& \frac{1}{4} \int \frac{d^2P d^2X}{(2\pi)^2} \frac{[\epsilon(\v P) - \mu]}{d^3} \Omega_z^\pm (2 + R \partial_R) \left (\frac{\vert \Delta (R) \vert}{R}\right)^2 \notag \\
&=&- \nu_v\int \frac{d^2P d^2X}{(2\pi)^2} \frac{\Omega_z}{R} \partial_R \frac{[\epsilon(\v P) - \mu]}{2d} \label{eq:tildeNzintermediate} \\
&=&2\pi \nu_v\int \frac{d^2P}{(2\pi)^2} \Omega_z [ n(R = \infty, \v P) -  n(R = 0, \v P)]
\end{eqnarray}
\end{subequations}
Here we introduce the semiclassical, electronic occupation $n(\v R, \v P) = 1/2 - {[\epsilon(\v P) - \mu]}/{2 d(\v R, \v P)}$ 
(the Bogoliubov angle is $\cos[\theta(\v R, \v P)] = 1-2 n (\v R, \v P)$).
We may now reinstall band (helicity) indices $\xi$ ($\pm$) for the total representation of the index
\begin{subequations}
\begin{align}
N_\pm(k_z) &= \nu_v [\sigma_{xy,\pm} (R = \infty) - \sigma_{xy,\pm} (R = 0)] \\
\sigma_{xy,\pm} (R) &= \sum_\xi \int \frac{d^2P}{2\pi} \Omega_{\xi, \pm,z}  n_\xi (R, \v P). \label{eq:IndexAppendix:sigma}
\end{align}
\label{eq:IndexAppendix}
\end{subequations}

\subsubsection{Boundary conditions}
The result Eq.~\eqref{eq:IndexAppendix} relates the index $N(k_z)$ to the difference of (spin) Hall conductivities at infinity and at zero. As such, it directly compares states with different topology to each other.
As explained in the main text, a Dirac semimetal is topological for $k_z \in (- k_z^*, k_z^*)$ and trivial otherwise. Therefore, $\sigma_{xy,\pm}(R = 0)$ displays topological quantization for an extended interval of $k_z$ and a topological transition occurs at $\pm k_z^*$.

In contrast, in the superconducting state, the system is gapped for all momenta and there is no topological transition as a function of $k_z$. In this case, Eq.~\eqref{eq:IndexAppendix} does not yield quantized response due to the nonuniversal behavior of $n_\xi (R, \v P)$ as a function of $\vert \Delta  (R)\vert$. However, we observe that $\sigma_{xy,\pm}(R = \infty)  = 0$ when $\vert \Delta  (R = \infty) \vert \rightarrow \infty$.
% because in that case $n_\xi (R = \infty, \v P) = 1/2$ and the total Chern number summed over all bands vanishes.

In fact, the Dirac superconductor is adiabatically connected to the superconducting state of trivial FeSC compounds (without inversion of $p$ and $d$ bands) and as such to vacuum. To illustrate this assertion in Fig.~\ref{fig:FiniteSystem}, we introduce a parameter $t_z$ to denote the strength of $z-$hopping. It is defined such that $t_z = t_0$ represents the band structure as in Fig.~\ref{fig:Dispersions} of the main text while $t_z = 0$ encodes a topologically trivial material without dispersion in $z$ direction.
We employ the following formal three step procedure: First, Fig.~\ref{fig:FiniteSystem} a)-c), one adiabatically increases $\vert \Delta\vert$ to a value $\Delta_{\rm help} \gg t_0$. As a next step one may adiabatically decrease $t_z$ from $t_0$ to $0$, Fig.~\ref{fig:FiniteSystem} d). Finally, one slowly reduces $\vert \Delta \vert$ from $\Delta_{\rm help}$ to zero, Fig.~\ref{fig:FiniteSystem} e). For a fully gapped s-wave superconductor the spectrum never closes for any $k_z$, hence the system is adiabatically connected to the topological trivial state and thus to vacuum. 

One may use this series of deformations to prove that in a finite system $N_\pm(k_z) = - \nu_v \sigma_{xy,\pm} (R=0)$, i.e. that the contribution from $R = \infty$ vanishes from Eq.~\eqref{eq:IndexAppendix}. In an isotropic system, $(t_z(R), \vert\Delta(R)\vert)$ traces a curve in the $(t_z, \vert \Delta \vert)$ plane. There are three regimes as a function of the radial coordinate, see the green solid curve in Fig.~\ref{fig:FiniteSystem} f): (i) normal state vortex core $(t_z(R), \vert\Delta(R)\vert) \vert _{R \ll \xi} \sim (t_0,0)$, (ii) bulk superconductor $(t_z(R), \vert\Delta(R)\vert) \vert _{\xi \ll R \ll L} \sim (t_0,\Delta_{\rm bulk})$, (iii) vacuum $(t_z(R), \vert\Delta(R)\vert) \vert _{L \ll R} \sim (0,0)$. The semiclassical procedure exposed in this appendix is capable of treating any vertical curves in the $(t_z, \vert \Delta \vert)$ plane. Horizontal curves result in $R$ dependence of the wave functions (and thus of $\mathcal A$, $\Omega_z$). An appropriate treatment would yield additional derivative terms in various places of our calculation, e.g. Eqs.~\eqref{eq:MoyalGeneralizedWT}, \eqref{eq:tildeNzintermediate}. However, as $\vert \Delta \vert \rightarrow \infty$ terms including $\partial_R \mathcal  A, \partial_R \Omega_z$ vanish from the final result for $N(k_z)$. This follows from the direct evaluation of Eq.~\eqref{eq:tildeNzintermediate} (we also checked this statement for terms featuring $\partial_R \mathcal  A$). 

Our formalism is thus capable to evaluate $N(k_z)$ for a trajectory $(t_z(R), \vert\Delta(R)\vert)$ which follows the blue dashed contour of Fig.~\ref{fig:FiniteSystem}. We obtain $N_\pm(k_z) = - \nu_v \sigma_{xy,\pm} (R=0)$, because the Berry curvature vanishes in the trivial system at $R = \infty$. As long as no singularities are being crossed, this result should hold for any continuously deformed integration contour. Since there are no singularities apart from the topological transition point $(t_z, \vert \Delta \vert) = (t_c, 0)$, we argue that $N_\pm(k_z) = - \nu_v \sigma_{xy,\pm} (R=0)$ for a system with physical boundary conditions which we represent by the green solid curve in Fig.~\ref{fig:FiniteSystem}. This concludes the derivation of Eq.~\eqref{eq:IndexSemiclassical} of the main text.

%%%%%%%%%%%%%%%%%
\begin{figure}
\includegraphics[scale=.3]{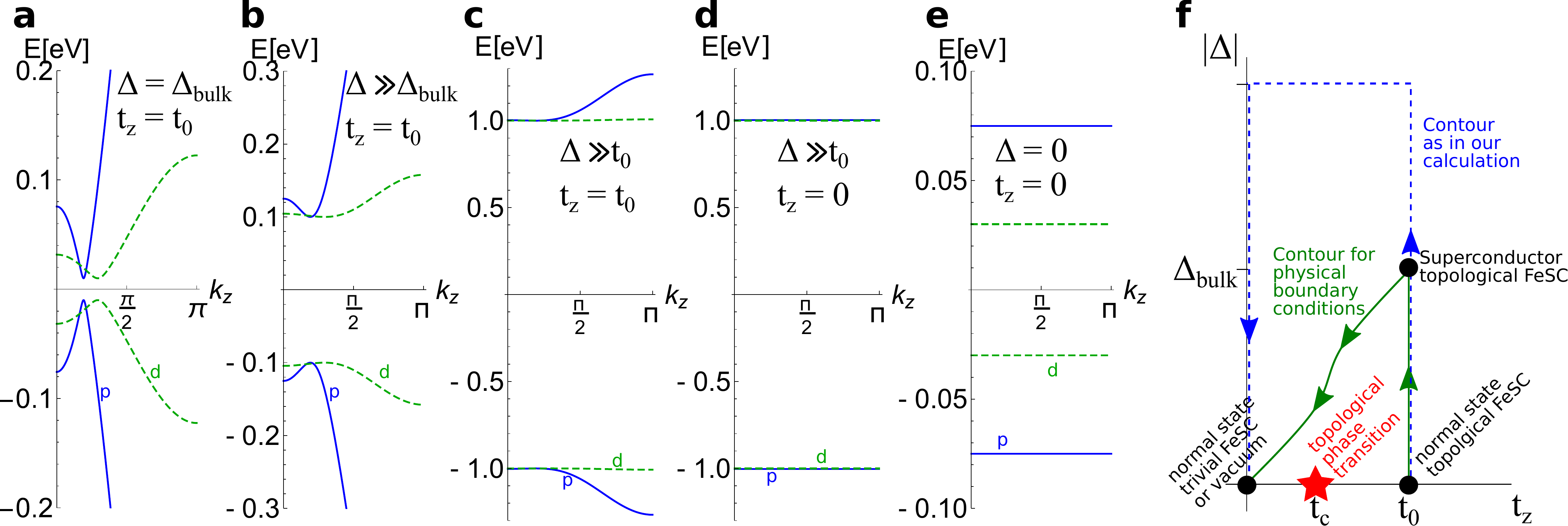}
\caption{Adiabatic deformation of the Dirac superconducting state into a trivial material. Panels a) - e): Bogoliubov spectra keeping the $j_z = \pm 3/2$ and $p$ bands of Fig.~\ref{fig:Dispersions} of the main text (energy relative to the Fermi level). The choice of parameters $t_z, \vert \Delta \vert$ is presented in the inset. f) Contour integration entering $N(k_z)$. Our calculation demonstrates the $N(k_z) = - \nu_v \sigma_{xy,\pm} (R=0) $ using the blue dashed contour. Since the only singularity of the spectrum resides at $(t_z, \vert \Delta \vert) = (t_c, 0)$, we conclude that the blue integration contour may be deformed into the green solid integration contour without changing the final result.}
\label{fig:FiniteSystem}
\end{figure}
%%%%%%%%%%%%%%%%%

\section{Quasiclassical calculation}

Here, we derive our results using the method summarized in Chapter 23.2 of Ref~\cite{VolovikBook} of the main text (i.e. a quasiclassical derivation of solutions). Again, we concentrate on a given Weyl sector with normal state Hamiltonian $H_{\v p} = v(p_x \sigma_x + p_y \sigma_y + p_z(k_z) \sigma_z) + M(p_z)$ (near the Weyl node $p_z$ is an odd function of $k_z$).
We denote $(p_x,p_y) = p_\perp (\cos(\theta), \sin(\theta)); (x,y) = \rho (\cos(\phi), \sin(\phi))$, and follow Ref.~\cite{VolovikBook} by transforming spatial coordinates to $s = \rho \cos(\phi - \theta)$ (position along a quasiparticle trajectory) and $b = \rho \sin(\phi - \theta)$ (impact parameter). We project the Hamiltonian onto the conduction band, and expand $\braket{u_{\v p} \vert \Delta(\v r) \vert u_{\v p}} \simeq \Delta(\v r) + \nabla_{\v r} \Delta(\v r) \cdot ( i  \nabla_{\v p_\perp} + \boldsymbol{\mathcal A})$ (in this Section we suppress indices $\xi$ and $\pm$). We exploit that $b$ is conserved (we here consider only the lowest energy state $b = 0$). In this notation we obtain
\begin{equation}
\mathcal H = \left (\begin{array}{cc}
- i v({p_\perp}) \partial_s & e^{i \theta} \left [ \vert \Delta \vert \text{sgn}(s) + \vert \Delta \vert ' (i \partial_{p_\perp} + \mathcal A_{\hat p_\perp}) + i \frac{\vert \Delta \vert}{{p_\perp} \vert s \vert} ( i \partial_\theta + {p_\perp} \mathcal A_\theta) \right] \\ 
e^{- i \theta} \left [ \vert \Delta \vert \text{sgn}(s) + \vert \Delta \vert ' ( i \partial_{p_\perp} + \mathcal A_{\hat p_\perp}) - i \frac{\vert \Delta \vert}{{p_\perp} \vert s \vert} ( i \partial_\theta + {p_\perp} \mathcal A_\theta) \right] & i v({p_\perp}) \partial_s
\end{array} \right ).
\end{equation}
The velocity is $v = v({p_\perp}) = \partial \epsilon({p_\perp},p_z)/\partial {p_\perp}$ and we suppressed the spatial dependence $\vert \Delta \vert = \vert \Delta (s) \vert$.
Contrary to topologically trivial systems, $\Omega_z = -p_z/[2\sqrt{p_z^2 + p_\perp^2}^3]$ (for $\mu >0$) does not vanish.  We henceforth project to the Fermi surface $p_\perp \rightarrow \sqrt{p_F^2 - p_z^2} \equiv q$ and we choose radial gauge, in which $\boldsymbol{\mathcal A} = \mathcal A_{\hat p_\perp} (\cos(\theta), \sin(\theta)) + \mathcal A_{\theta} (-\sin(\theta),\cos(\theta))$ with $\mathcal A_{\hat p_\perp} = 0$ and 
\begin{equation}
\mathcal A_\theta (p_\perp) = - \frac{\text{sgn}(p_z)}{2 \sqrt{p_\perp^2 + p_z^2}} \frac{p_\perp}{\sqrt{p_\perp^2 + p_z^2} + \vert p_z \vert} \simeq -\frac{\text{sgn}(p_z)}{2q} \frac{1 - p_z^2/p_F^2}{1 + \vert p_z \vert / p_F}.
\end{equation}
Using the standard unitary transformation $e^{i \theta \tau_z/2}$ we obtain in close analogy to Ref.~\cite{VolovikBook}
\begin{equation}
\tilde{\mathcal H} = \left ( \begin{array}{cc}
- i v \partial_s & \vert \Delta(s) \vert \text{sgn}(s) + i \frac{\vert \Delta \vert}{q \vert s \vert} \left (i \partial_\theta - \frac{1}{2} + q \mathcal A_\theta(q)\right) \\ 
 \vert \Delta(s) \vert \text{sgn}(s) - i \frac{\vert \Delta \vert}{q \vert s \vert} \left (i \partial_\theta + \frac{1}{2} + q \mathcal A_\theta(q)\right) & i v  \partial_s
\end{array} \right). \label{eq:HtildeVolovik}
\end{equation}
In view of the hierarchy of scales $q\vert s \vert \sim \sqrt{p_F^2- p_z^2} \xi \gg 1$ we follow Eq.~(23.16) of~\cite{VolovikBook} in omitting the second term of the off-diagonal elements. This leads to standard bound states in the vortex core. The perturbative inclusion of terms order $\omega_0 \sim \Delta/(q \xi)$ implies a low energy Hamiltonian $H = \omega_0 [Q  - q \mathcal A_\theta]$ with $Q = -i \partial_\theta$ the angular momentum operator with usual Caroli-deGennes-Matricon quantization $Q = n + 1/2$, cf. Eq.~(23.23) of \cite{VolovikBook}. The gauge-symmetry enforced shift $q \mathcal A_\theta$ also follows directly from Eq.~\eqref{eq:HtildeVolovik}. Since $q \mathcal A_\theta \simeq - \text{sgn}(p_z)/2 - p_z/(2p_F)$ near $p_z =0$ we readily find chiral, dispersive zero modes at the projection of the Weyl node with velocity $ \vert \Delta \vert/(p_F^2 \xi) \sim v \vert \Delta^2 \vert/\mu^2$. The inclusion of the second helical sector then implies two counterpropagating helical Majorana modes in accordance with the quantum mechanical calculation exposed above.

\subsection{Comparison to $^3$He-A and $^3$He-B}

Using the technique employed in this section, one may readily compare to other systems studied with the same technique \cite{Volovik2011,VolovikBook}. (1) Trivial s-wave superfluids: In view of the trivial normal state band structure, $q \mathcal A_\theta$ is absent and the spectrum is gapped, $E = \omega_0 (n+1/2)$. (2) Two-dimensional $p+ip$ superconductor $H = (\v p^2/2m - \mu)\tau_z + \Delta (p_x \tau_x + p_y \tau_y)$: The trivial normal state Hamiltonian again implies $q \mathcal A_\theta = 0$. However, the winding of the gap function in momentum space adds an additional shift of 1/2 in the low energy Hamiltonian $H = \omega_0 [Q - 1/2]$, which implies a gapless spectrum $E = \omega_0 n$ (the $n = 0$ state is usually called Majorana mode).
(3) Vortex-disgyration ($\mathbb Z_2$ vortex) in $^3$He-A: For each spin $\sigma = \pm 1$, this phase can be described by $H = (\v p^2/2m - \mu)\tau_z + \sigma \Delta (p_x \tau_x + p_y \tau_y)$. Therefore, the conclusions of point (2) imply a non-dispersive, doubly degenerate flat band of Majoranas in a finite interval of $p_z$, as long as spin components do not mix. (4) Symmetric vortex in $^3$He-B, where $H = (\v p^2/2m - \mu)\tau_z + \Delta  \tau_x \v p \cdot \boldsymbol \sigma$: Again, $q \mathcal A_\theta = 0$, and the non-trivial $\v p$ dependence implies $H = \omega_0 [Q - 1/2]$. However, the spin-orbit coupled gap structure lifts the two-fold degeneracy of the one-dimensional spectrum except for helical touching points.

\newpage
\end{widetext}


\begin{thebibliography}{59}
\bibitem [{Foo()}]{FootnoteFS}%
  \BibitemOpen
  \href@noop {} {}\bibinfo {note} {Strictly speaking, if $k_z$ states are
  filled up to a finite Fermi momentum $k_F$ around the Weyl node, quantization
  occurs for the smaller interval $k_z \in (-k_z^*+k_F,
  k_z^*-k_F)$.}\BibitemShut {Stop}%
\end{thebibliography}
\end{document}